\documentclass[a4paper,12pt]{article}
\usepackage{epsf}
\newcommand{\bmat}{\left(\begin{array}}
\newcommand{\emat}{\end{array}\right)}

\def\beq{\begin{equation}}
\def\eeq{\end{equation}}
\def\beqa{\begin{eqnarray}}
\def\eeqa{\end{eqnarray}}

\def\-{\hphantom{-}}

\def\s2{\frac{1}{\sqrt2}}

\def\beq{\begin{equation}}
\def\eeq{\end{equation}}
\def\beqa{\begin{eqnarray}}
\def\eeqa{\end{eqnarray}}
\def\ba{\begin{array}}
\def\ea{\end{array}}

\def\IF{\relax{\rm I\kern-.18em F}}
\def\II{\relax{\rm I\kern-.18em I}}
\def\IP{\relax{\rm I\kern-.18em P}}
\def\IC{\relax\hbox{\kern.25em$\inbar\kern-.3em{\rm C}}}
\def\IR{\relax{\rm I\kern-.18em R}}

\def\cp{{\cal P}}

\def\Dsl{\,\raise.15ex\hbox{/}\mkern-13.5mu D} %this one can be subscripted
\def\IZ{Z\kern-.4em  Z}

 \def\cp#1{\relax\ifmmode {\IP\kern-2pt{}_{#1}}\else $\IP\kern-2pt{}_{#1}$\=fi}
\catcode`\@=11
\newdimen\@rotdimen
\newbox\@rotbox

\def\@vspec#1{\special{ps:#1}}%  passes #1 verbatim to the output
\def\@rotstart#1{\@vspec{gsave currentpoint currentpoint translate
   #1 neg exch neg exch translate}}% #1 can be any origin-fixing transformation
\def\@rotfinish{\@vspec{currentpoint grestore moveto}}% gets back in synch
\def\@rotr#1{\@rotdimen=\ht#1\advance\@rotdimen by\dp#1%
   \hbox to\@rotdimen{\hskip\ht#1\vbox to\wd#1{\@rotstart{90 rotate}%
   \box#1\vss}\hss}\@rotfinish}
\def\@rotl#1{\@rotdimen=\ht#1\advance\@rotdimen by\dp#1%
   \hbox to\@rotdimen{\vbox to\wd#1{\vskip\wd#1\@rotstart{270 rotate}%
   \box#1\vss}\hss}\@rotfinish}%
\def\@rotu#1{\@rotdimen=\ht#1\advance\@rotdimen by\dp#1%
   \hbox to\wd#1{\hskip\wd#1\vbox to\@rotdimen{\vskip\@rotdimen
   \@rotstart{-1 dup scale}\box#1\vss}\hss}\@rotfinish}%
\def\@rotf#1{\hbox to\wd#1{\hskip\wd#1\@rotstart{-1 1 scale}%
   \box#1\hss}\@rotfinish}%
\def\rotate{\@ifnextchar[{\@rotate}{\@rotate[l]}}
\def\@rotate[#1]#2{\setbox\@rotbox=\hbox{#2}\@nameuse{@rot#1}\@rotbox}

\catcode`\@=12
\begin{document}

%----------------------------------------------------------------------%
%  numbering equations with section number
%----------------------------------------------------------------------%
\makeatletter \@addtoreset{equation}{section} \makeatother
\renewcommand{\theequation}{\thesection.\arabic{equation}}
%----------------------------------------------------------------------%
%  title page
%----------------------------------------------------------------------%
\pagestyle{empty}
%----------------------------------------------------------------------%
%  Resetting of counters
%----------------------------------------------------------------------%
%\setcounter{page}{0}
\pagestyle{empty}
%\vspace{0.5in}
%\rightline{FTUAM-03-09}
%\rightline{IFT-UAM/CSIC-03-17}
\rightline{\today}
\vspace{3.0cm}
\setcounter{footnote}{0}

%\begin{center}
%\underline{\LARGE{****  DRAFT VERSION *****}}
%\end{center}

\begin{center}
\LARGE{\bf Nambu-Lie 3-Algebras on Fuzzy 3-Manifolds }
\\[8mm]
%\medskip
{\Large{ Minos Axenides$^{1}$ and Emmanuel Floratos$^{1,2}$}}
% and Christos Kokorelis$^{1}$}}
\\[6mm]
 \normalsize{\em $^1$ Institute of Nuclear Physics, N.C.S.R. Demokritos,
GR-15310, Athens, Greece}\\
\normalsize{\em $^2$  Department of Physics, Univ. of Athens,
GR-15771 Athens, Greece}\\[5mm]  axenides@inp.demokritos.gr; \ \ \   mflorato@phys.uoa.gr 
\end{center}
\vspace{6.0mm}
%%%%%%%%%%%%%%%%%%%%%%%%%%%%%%%%%%%%%%%
%\begin{center}
%\begin{minipage}[h]{14.5cm}
\begin{center}
{\large {\bf Abstract}}
\end{center}

\vspace{5.0mm}

We consider Nambu-Poisson 3-algebras on three dimensional manifolds  $ {\cal M}_{3} $, 
such as the  Euclidean 3-space  $R^{3}$, the 3-sphere  $S^{3}$  as well as 
the 3-torus  $T^{3}$. We demonstrate that in the Clebsch-Monge gauge, the Lie 
algebra of volume preserving diffeomorphisms  $SDiff({\cal M}_{3})$  is identical to the
Nambu-Poisson algebra on  ${\cal M}_{3}$. Moreover the fundamental identity 
for the Nambu 3-bracket 
is just the commutation relation of  $ SDiff({\cal M}_{3})$. We propose a quantization 
prescription for the Nambu-Poisson algebra which provides us with the correct 
classical limit. As such it possesses all of the expected classical properties 
constituting, in effect, a concrete representation of Nambu-Lie 3-algebras.

\newpage

%----------------------------------------------------------------------%
%  Resetting of counters
%----------------------------------------------------------------------%
\setcounter{page}{1} \pagestyle{plain}
\renewcommand{\thefootnote}{\arabic{footnote}}
\setcounter{footnote}{0}
%----------------------------------------------------------------------%
%  Paper begins
%----------------------------------------------------------------------%
\tableofcontents
\section{Introduction}

Recently the Nambu-Lie (NL) 3-algebras\cite{Nam,Fil,Tak} have been in the focus of 
interest since they appear as gauge symmetries of new superconformal Chern-Simons
non-abelian theories in $2+1$  dimensions with the maximun allowed number of 
$N=8$ linear supersymmetries. \cite{JS,BH,BL,Gust}. These theories explore the low 
energy dynamics of the microscopic degrees of freedom of
coincident ${\cal M}_{2}$  branes  and constitute the boundary conformal 
field theories of the bulk 
$ AdS_{4} \times S_{7} $  exact 11-dimensional supergravity  backgrounds of supermembranes 
\cite{HLW}. These mysterious new symmetries, the NL 
3-algebras represent the implementation of non-associative algebras of coordinates
of charged tensionless strings, the boundaries of open $ {\cal M}_{2}$ branes in 
antisymmetric field magnetic backgrounds of $ {\cal M}_{5}$ branes in the $ {\cal M}_{2}-
{\cal M}_{5}$ system \cite{Ber}. The NL 3-algebras are either operator or matrix 
representation of the classical Nambu-Poisson (NP) symmetries of world volume preserving
diffeomorphisms of ${\cal M}_{2}$ branes \cite{BSTT}. 
Indeed at the classical level the supermembrane
Lagrangian, in the covariant formulation, has the world volume preserving 
diffeomorphism symmetry 
$SDiff[M_{2+1}]$. The Bagger-Lambert-Gustaffson  3-algebras presumably 
correspond to the quantization
of the rigid motions in this infinite dimensional group, 
which describe the low energy excitation
spectrum of the $M_{2}$ branes \cite{Tow}.

In the light-cone (LC) gauge, the membrane symmetries
reduce to the area preserving diffeomorphisms of the membrane surface and the matrix 
truncation of this infinite dimensional  group by $SU[N]$ \cite{Hop,WHN} 
provided a basic ingredient
for the Matrix-Model proposal \cite{BFSS}.

In ref.\cite{Flo} the SU(N) truncation, was interpreted in terms of the matrix algebra 
of finite quantum mechanics on a discretized membrane surface(discrete non 
commutative phase space).
So one naturally could ponder about the existence of a discretized membrane 
world volume of $2+1$
dimensions and a Matrix model on it as the finite quantum mechanics in three dimensions.
Three dimensional classical phase spaces may arise in Nambu mechanics 
\cite{Nam} with the ensuing subtle issues of its quantization \cite{Tak,CZ,ALMY,Hop2,MT}.

Apart from the ${\cal M}_{2}$ brane dynamics, the 3-d volume preserving diffeomorphism group 
appears as the basic symmetry also in  the LC gauge Hamiltonian of 
$p=3$ superbranes \cite{BSTT}, where all the interaction terms are expressed in terms 
 of the Nambu 3-bracket. 

In ref.\cite{AF} we exploited the NP 3-algebras to find explicit rotating, 
rigid body (lowest energy),   solutions of LC   $S^3$ and $T^3$ branes in toroidally 
compactified
higher dimensional flat spaces. 
A Matrix Model analog of these solutions 
and more generally
for the LC dynamics for $p=3$  branes is lacking.
We would like to notice at this point
the Matrix model that ref.\cite{Jab} has proposed under the name 
"Tiny Graviton Matrix Model" for spherical (fuzzy $ S^3 $)  $D_{3}$ branes, 
as well as the construction of fuzzy $ S^{3} $ spheres \cite{GRBC}. 

A completely new and radical approach has been taken by the advocates of 
cubic matrix algebras
which presumably discretize consistently three dimensional manifolds 
in a similar way that usual two-index
matrices discretize surfaces. 
This direction is interesting by itself but the difficulties 
seem to be both intriguing and challenging at the same time \cite{Mats,Kaw}.

The most mathematically complete quantization scheme for the Nambu 3-bracket 
up to now is by ref.\cite{SDT} where an algebraic topological quantization, the Zariski
 $*$ quantization and variations thereof, has been proposed, but the algebraic complexity 
 of the scheme seems to hide important physical and geometrical aspects of the problem.
 All the other present proposals are violating, in general, the basic properties of the 
3-bracket such as Leibniz 
and the Fundamental Identity \cite{Tak}. 
For a critical and rather complete discussion of the state of art
we refer to ref.\cite{CZ} and for general perspectives of the 
quantization of Nambu mechanics see ref.\cite{MT}.

In this work we will exploit the relation of the classical Nambu-Poisson algebra 
in euclidean 3-d spaces ( ${\cal M}_{3} = R^{3} , S^{3} , T^{3} $,
or 3-manifolds embeddable
in $ R^{4} $ ) with the volume
preserving diffeomorphism algebras $ SDiff({\cal M}_{3})$. 
Moreover we shall propose
a consistent quantization prescription which offers a concrete realization of the 
Nambu-Lie 3-algebras on these spaces. 

In section 2 we are going to review the problem of quantization of Nambu mechanics.

In section 3 we shall discuss the basic properties of NP 3-algebras which 
correspond to particular cases  of 3-manifolds and
pertain to Nambu mechanics. 

In section 4 we will present the Lie algebra of volume preserving diffeomorphisms 
$ SDiff(R^{3}) $ in the Clebsch-Monge gauge, their relation with the NP
3-algebras on $ R^{3} $ as well as on $ T^{3}$  and Nambu mechanics, 
which can be represented as flow
equations of incompressible fluids. 

In section 5 we will discuss the role of Clebsch-Monge gauge in the case 
of a non-trivial topology which is present in Nambu flows with vortices.

In section 6 we will propose a new quantization scheme for 
the Nambu mechanics  which posseses naturally the correct classical limit. 

Finally in section 7 we quantize particular Nambu-Poisson 3-algebras  consistently 
with the classical properties of  a) complete antisymmetry
 b) Leibniz and c) Fundamental Identity .  

The proposed quantization prescription is based on the intuitive idea that at each point 
of a 3-space the volume element ( Nambu 3-bracket) 
is defined by a triple family of coordinate
surfaces. In an analogous way the quantum volume element should be defined by a triple 
family of intesecting fuzzy coordinate  surfaces. 
The resulting quantum 3-algebras provide concrete realizations of 
Nambu-Lie 3-algebras.

%------------------------------------------------------------------------------------------

\section{On Classical Nambu Dynamics in 3-D Phase Space and its Quantization}

%-----------------------------------------------------------------------------------------

Nambu in his classic paper \cite{Nam} introduced new dynamical systems
with arbitrary even or odd dimensions of "phase space"  possessing as 
fundamental symmetries 
the volume preserving diffeomorphism group in the place of symplectic 
diffeomorphisms \cite{Tak,CZ,Nam2}. 
The new equations of motion in the phase space 
$ M\equiv R^{n} $ are analogous to Hamilton-Poisson equations as follows:

\beq
\frac{d x^{i}}{d t } = \{ x^{i} , H_{1} , \cdots , H_{n-1} \} \ ,
\eeq
where the n-bracket is defined as : 

\beq
\{ f_{1} , \cdots , f_{n} \} = \epsilon^{i_{1} \cdots i_{n} } \partial^{i_{1}}f_{1}
 \partial^{i_{2}}f_{2} \cdots  \partial^{i_{n}}f_{n} \ ,
\eeq
for any functions $ f_{1}, \cdots , f_{n} \in C^{\infty}(R^{n}) $ and 
$ i_{1}, \cdots , i_{n} = 1,\cdots , n .$ 

The n-1 "Hamiltonians" $ H_{1}, \cdots , H_{n-1} $ determine the phase-space
trajectory in a geometrical way. There is also a corresponding Liouville equation
for any observable $ f \in C^{\infty}(R^{n}) $ ,

\beq
 \frac{d f}{d t } = \partial^{i} f \cdot \dot {x}^{i} = 
\{ f , H_{1} , \cdots , H_{n-1} \} \ .
\eeq
The n-1 Hamiltonians are conserved in  time. Given the initial position in the 
phase-space $ x^{i}_{0}= x^{i}(t=0) $ they take the values

\beq
h^{i} = H_{i} (x_{0}) \ \ ; \ \ \ \ \ \ \ \ \ \ i=1,2,\cdots , n-1 \ .
\eeq 
The intersection of hypersurfaces

\beq
H_{i} (x) = h^{i} \ \ ; \ \ \ \ \ \ \ i=1,\cdots , n-1 \ ,
\eeq
gives the geometrical shape of the trajectory passing through the point 
$ x_{0} \in R^{n} $ \cite{CZ}. This is the reason why the Nambu 3-d dynamical system
is regarded as a toy model for completely integrable systems. 
The basic properties of the n-bracket are :

1) Linearity

\beq
\{ \alpha f_{1} + \beta g_{1} , f_{2} , \cdots , f_{n} \} = \alpha \{ f_{1} , f_{2} ,
\cdots , f_{n} \} + \beta \{ g_{1} , f_{2} , \cdots , f_{n} \} \ . 
\eeq 

2) Antisymmetry

\beq
\{ f_{\sigma(1)} ,\cdots , f_{\sigma(n)} \} = ( -1 )^{\sigma} 
\{ f_{1} , \cdots , f_{n} \} \ , \sigma \in S_{n}. 
\eeq

3) Leibniz Rule

\beq
\{ f \cdot g , f_{1} , \cdots , f_{n} \} = f \{ g , f_{2} , \cdots , f_{n} \} +
\{ f , f_{2} , \cdots , f_{n} \} g \ .
\eeq
To the above we must finally add an extension of the Jacobi identity for the Poisson
brackets, i.e. the Fundamental Idcentity(FI)

\beqa
\{ \{ f_{1} , \cdots , f_{n} \} , f_{n+1} , \cdots , f_{2n-1} \} &=& 
\{ \{ f_{1} , f_{n+1} , \cdots , f_{2n-1} \} , f_{2} , \cdots , f_{n} \} + \nonumber \\
&+& \{ f_{1} , \{ f_{2} , f_{n+1} , \cdots , f_{2n-1} \} , f_{3} , \cdots , f_{n} \} 
\nonumber \\
&+& \cdots + \{ f_{1} , \cdots , f_{n-1} , \{ f_{n} , f_{n+1} , \cdots , f_{2n-1} \} \ ,
\eeqa
for $ ( f_{i} )_{i=1,2,\cdots , 2n-1} \in C^{\infty} ( R^{n} ) \ . $ 

The FI can be proved directly either through the use of the definition of the n-bracket 
or by following up the time evolution of the observable $ \{ f_{1} , \cdots , f_{n} \} $
on the phase - space trajectories with respect to the Hamiltonian 
$ H_{1} = f_{n+1} , \cdots , H_{n-1} = f_{2n-1} $. This identity gurrantees 
the fact that if $ ( f_{i} )_{i=1 , \cdots , n} $ are each seperately conserved 
quantities, then the observable $ \{ f_{1} , \cdots , f_{n} \} $ is also conserved. 
It is this property
that becomes an obstacle to the quantization of Nambu Dynamics. We would like to have
a Heisenberg-Nambu extensions of the Heisenberg quantum mechanical eqs :

\beq
i\hbar \frac{d \hat{x}^{i}}{d t } \ = \ [ \hat{x}^{i}, \hat{H}_{1}, \cdots ,
 \hat{H}_{n-1} ] ,
\eeq
where we pass from the classical position vector  
$ ( x^{i} )_{i=1 , \cdots , n} $, classical "Hamiltonian" 
$ ( H_{i} )_{i=1,\cdots, n-1 } $
and  the Nambu-Poisson n-bracket(2.2) to their corresponding  quantum operator versions 
$ (\hat{x}^{i})_{i=1,\cdots ,n} , (\hat{H}_{i})_{i=1,\cdots ,n-1} $, and 
Nambu-Lie n-commutator  (2.10) \cite{Nam,Fil}. 
All proposals to date for the n-commutator or the Quantum Nambu bracket
fail,  in general, to satisfy both the Leibniz rule and the FI,  
which are crucial for the consistency
of the time evolution (2.10). 
It is also significant that most of them, also fail to reproduce the
correct classical limit.
In ref.\cite{CZ} there is a detailed discussion of the problem along with a  
specific resolution through the adoption of different time evolutions for different
superselection sectors of the Hilbert space. 

Nambu proposed to abandon the Leibniz property  and the FI (i.e. to abandon consistency
with unique time evolution, Liouville eqn.) and insist on the linearity 
and antisymmetry properties.
More specifically for any n operators $ ( \hat{F}_{i})_{i=1,2,\cdots ,n } $  he proposed
the definition 

\beq
[ \hat{F}_{1} , \cdots , \hat{F}_{n} ] = \ \ \  \sum_{\sigma \in S_{n}} (-1)^{\sigma} 
\hat{F}_{\sigma_{1}} \cdots \hat{F}_{\sigma_{n}} .
\eeq
For even $ n=2,4,\cdots ,$ there are interesting identities which reduce the RHS of 
eq.(2.11) into products of all commutator pairings 
$ [ \hat{F}_{\sigma_{n}} , \hat{F}_{\sigma_{m}} ] $,
which qurantee the correct classical limit.  For odd values
$ n=3, 5, \cdots , $ this property does not hold. 
One way to go, is to adopt an "odd-even"  reduction
through the use of fixed operators $ \hat{F}_{0} $ and define :

\beq
[ \hat{F}_{1} , \cdots , \hat{F}_{2k+1} ] \ \ \  = \ \ \ [ \hat{F}_{0} , 
\hat{F}_{1} , \cdots , \hat{F}_{2k+1} ] .
\eeq  
In the next section  we present explicit constructions of Nambu-Poisson 
algebras for the case $ n=3 $, i.e. for three dimensional manifolds and especially for 
$ R^{3} , S^{3} , T^{3} $ as well as for 3-d manifolds embeddable in $R^4$
by level set Morse functions.

%------------------------------------------------------------------------------------------

\section{Nambu-Poisson 3-Algebras}

%------------------------------------------------------------------------------------------

Nambu-Poisson (NP) algebras have been introduced in ref.\cite{Tak}. 
We consider generalized Nambu 3-brackets on $n$ dimensional
manifolds $M_{n}$  which are defined through a 3-index
antisymmetric tensor field $ \omega^{ijk}(x) $ for 
$ x \in M_{n}, i,j,k=1,2, \cdots, n $

\beq
\{ f , g , h \} = \omega^{ijk}(x) \partial^{i} f \partial^{j} g \partial^{k} k .
\eeq
We observe that linearity, antisymmetry and the Leibniz rule are satisfied by definition. 
We shall impose further the Fundamental Identity on the tensor field $\omega $. 
For $ f=x^{i},  \ \ g =x^{j}, \ \ h = x^{k} $ we have the Nambu-Poisson 3-algebras 
for the  coordinates

\beq
\{ x^{i} , x^{j} , x^{k} \} = \omega^{ijk}(x) \ \ \ ; \ \ \ \ \ \ \ \ \ \ \ i,j,k=1,\cdots ,n \ .
\eeq
The FI  imposed on the coordinate functions is:

\beqa
\{ \{ x^{i} , x^{j} , x^{k} \} , x^{l} , x^{m} \} &=& \{ \{ x^{i} , x^{l} , x^{m} \} ,
x^{j} , x^{k} \} \nonumber \\ &+& \{ x^{i} , \{ x^{j} , x^{l} , x^{m} \} , x^{k} \} +
\{ x^{i} , x^{j} , \{ x^{k} , x^{l} , x^{m} \} \} ,
\eeqa 
or by using (3.1-2)

\beq
\omega^{plm} \partial^{p} \omega^{ijk} = \omega^{pjk} \partial^{p} \omega^{ilm} +
\omega^{ipk} \partial^{p} \omega^{jlm} + \omega^{ijk} \partial^{p} \omega^{klm}
\ \ ; \ \ \ \ \ p,i,j,k,l,m=1,2,\cdots ,n .
\eeq
For a smooth manifold $ {\cal M}_{n}$ 
of $ dim {\cal M } = n $, which is equiped with a non-degenerate 
3-form  $\omega $ and satisfies (3.4), it can be shown that this 
condition is too strong. In fact n  must be restricted to be $ n=3 $ \cite{Gau}. 
It is identified as "the indecomposability" condition for the Nambu 3-tensor 
$\omega $. As a further unexpected refinement  we can choose locally coordinates 
on the 3-manifold $ {\cal M }_{3} $ so that

\beq
\omega^{ijk} = \epsilon^{ijk} \ \ ;  \ \ \ \ \ \ \ i,j,k=1,2,3 
\eeq
is the $ R^{3} $ Nambu form. If  $ {\cal M }_{3} $ possesses a metric with a volume 
element $ \sqrt{g} $ then the typical form of the Nambu tensor  
takes the form

\beq
\{ x^{i} , x^{j} , x^{k} \} = \frac{ \epsilon^{ijk} }{\sqrt{g}} \ .
\eeq
Relation (3.5) is analogous to the existence of local coordinates in symplectic 
manifolds with $ \sqrt{g} = 1 $ \cite{Darb}. 

In order to construct non-trivial examples of  3-algebras  
we follow the crucial observation of L.Takhtajan that the Nambu n-brackets 
in $ R^{n} $ rel.(2.2) create a
tower of lower dimensional brackets of order $ n-1 , n-2 , \cdots $ on submanifolds
which are embedded in $ R^{n}$ . In order to be more specific, let us consider a smooth
3-Manifold $ {\cal M }_{3} $ embedded in $ R^{4} $ through a level-set function (Morse
function)  :

\beq
h(x^{1},x^{2} , x^{3} , x^{4} ) = c ,
\eeq
with $ c \in R $ fixed. 
Then by using the FI in $ R^{4} ( n=4 $ in rel. 2.9 ) we can
check that the 3-bracket on $ R^{4}$

\beq
\omega^{ijk} = \epsilon^{ijkl} \partial^{l} h \ \ ; \ \ \ \ \ \ \ \ i,j,k,l=1,2,3,4 
\eeq
satisfies the FI rel.(3.4) with $ n=4$. For example if h is a linear function

\beq
h(x^{1},x^{2} , x^{3} , x^{4} ) = \alpha^{i} x^{i} \ \ ; \ \ \ \ \ \ \ i=1,2,3,4 ,
\eeq
then we obtain the constant Nambu-Poisson(NP)  3-algebras:

\beq
\{ x^{i} , x^{j} , x^{k} \} = \epsilon^{ijkl} \alpha^{l} \ \ ; \ \ \ \ \  i,j,k,l= 1,2,3,4 .
\eeq
If h is a quadratic function, representing the sphere $ S^{3} \subset R^{4} $

\beq
h = \frac{1}{2}( x^{i})^{2},
\eeq
then we have the linear Nambu-Poisson 3-algebra

\beq
\{ x^{i} , x^{j} , x^{k} \}_{S^{3}} = \epsilon^{ijkl} x^{l}  \ \ ; \ \ \ \ \ \ 
 i,j,k,l= 1,2,3,4 \ .
\eeq
We observe that the most general NP 3-algebra rel.(3.8)

\beq
\{ x^{i} , x^{j} , x^{k} \}_{h} = \epsilon^{ijkl} \partial^{l} h \ \ ; \ \ \ \ \ \ \ 
i,j,k,l= 1,2,3,4 \ ,
\eeq
has h as Casimir

\beq
\{ x^{i} , x^{j} , h \}_{h} = 0 \ \ ; \ \ \ \ \ \ \ \ \ \ \ \ i,j = 1 2,3,4 \ ,
\eeq
and the 3-form $\omega^{ijk} $ (3.7) is thus degenerate and we bypass Gautheron's
 theorem \cite{Gau}

\beq
\omega^{ijk}\partial^{k} h = 0 \ \ ; \ \ \ \ \ \ \ \ \ \ \ \ i,j,k=1,2,3,4 .
\eeq
Restriction of the algebra (3.13)  on the surface (3.7)
gives a non-degenerate 3-form $ \omega^{ijk} , i,j,k=1,2,3 $ which satisfies the F.I..

Let us now proceed to present three examples of 3-algebras such as 
$R^{3}, S^{3} $ and 
$ T^{3} $ . 
Starting out with $ R^{3} $  the  3-algebra of coordinates is  :

\beq
\{ x^{i} , x^{j} , x^{k} \} = \epsilon^{ijk} \ \ ; \ \ \ \ \ \ \ \ i,j,k=1,2,3 .
\eeq
By using the Leibniz property we can write down the algebra for the monomial 
basis

\beq
x^{n} = x_{1}^{n_{1}} x_{2}^{n_{2}} x_{3}^{n_{3}} \ \ ; \ \ \ \ \ \ \ \ \ 
n_{1},n_{2},n_{3} = 0,1,2,\cdots \ ,
\eeq

\beq
\{ x^{n} , x^{m} , x^{l} \} = n \cdot (m \times l ) x^{n+m+l-(1,1,1)} .
\eeq
For the 3-torus $ T^{3} $ the algebra for the periodic function basis: 

\beq
e_{n} = e^{i n \cdot x} \ ,
\eeq
with $ n = (n_{1} , n_{2} , n_{3}) \in Z^{3}$   and 
$ x=(x_{1} , x_{2} , x_{3} ) \in ( 0 , 2\pi )^{3} $ 

is given by \cite{Hop2,AF,Jab,Mats}

\beq
\{ e_{n} , e_{m} , e_{l} \} = -i n\cdot (m \times l) e_{n+m+l} \ \ ; \ \ \ \ \ n,m,l \in Z^{3} .
\eeq 
For the case of a sphere $S^{3}$  \cite{AF,Jab}, rel.(3.12) 
we use polar coordinates to project on the surface :

\beqa
e^{4} &=& cos\theta_{3} \ \ \ \ \  \nonumber \\
e^{3} &=& cos\theta_{2} sin\theta_{3} \ \ \ \ \ \ \  \nonumber \\
e^{2} &=& sin\theta_{1} sin\theta_{2} sin\theta_{3} \ \ \ \ \  \\ 
e^{1} &=& cos\theta_{1} sin\theta_{2} sin\theta_{3} \ \ \ \ \ \ \nonumber \ ,
\eeqa
$ \theta_{1} \in (0,2\pi), \theta_{2}, \theta_{3} \in (0,\pi) $

\beq
\{ e^{i} , e^{j} , e^{k} \}_{S^{3}} = \frac{1}{sin^{2}\theta_{3}sin\theta_{2}} 
\epsilon^{qrs}\partial_{\theta_{q}} e^{i} \partial_{\theta_{r}} e^{j} 
\partial_{\theta_{s}} e^{k} = \epsilon^{ijkl} e^{l} \ .
\eeq
By using the Leibniz property  3-algebra on $S^{3}$ 
it is possible to write down explicitly,
for a basis of hyperspherical harmonics the corresponding NP $S^{3}$ 3-algebras,

\beq
 Y_{a}(\Omega) = Y_{nlm}(\theta_{3},\theta_{2},\theta_{1}) \ \ ; \ \ \ \ \ \ a=(nlm) , \ \ \ \ \ \ 
m=-l,\cdots , l \ \ , \ \ \ \  l=0,\cdots , n-1 \ \ ,
\eeq

\beq
\{  Y_{\alpha}, Y_{\beta}, Y_{\gamma} \} = f_{\alpha\beta \gamma}^{\delta} Y_{\delta},
\eeq
where $ f_{\alpha\beta \gamma}^{\delta} $ can be expressed in terms of 6j symbols 
of SU(2) ( $ O(4) \sim SU(2) \times SU(2) $). 
For volume preserving diffeomorphisms
of $S^{3}$ the usual commutators have been worked out with vector spherical harmonics
in ref.\cite{Dow}.  

In the rest of this section we shall apply the induction procedure to get a 
simpler geometrical meaning for the  3-brackets of the Nambu Dynamics 
in $R^{3} $ ( similarly for $T^{3}$ and/or $S^{3}$ ). In this case evolution eqs. are 
controlled by two Hamiltonians $H_{1}, H_{2} \in C^{\infty}(R^{3})$ and are given by

\beq
\frac{d x^{i}}{d t } = \{ H_{1}, H_{2} \}_{i} \ \ ; \ \ \ \ \ \ \ \ \ \  i=1,2,3 ,
\eeq
where the Poisson brackets $\{ H_{1} , H_{2} \}_{i}$ are:

\beq
\{H_{1}, H_{2} \}_{i} =  \epsilon^{ijk}\partial^{j} H_{1} \partial_{k}H_{2} 
\ \ ; \ \ \ \ \ \ \ \ \ \ \ i=1,2,3 .
\eeq 
Essentially we have three pairs of canonical variables 
$(x^{1} x^{2}) , (x^{2} x^{3}), (x^{3} x^{1})$  with coupled evolution eqs. 
It is possible to bring them into the usual Hamilton's eqn. as follows.
We choose one "Hamiltonian" say $ H_{2}$
to describe the geometry of a two dimensional phase-space embedded in $ R^{3} $ , 
$ H_{2}(x) = C $   and we write :

\beq
\frac{d x^{i}}{d t } = \{ x^{i} , H_{1} \}_{H_{2}} \ \ ; \ \ \ \ \ \ \ \ \ i=1,2,3 \ ,
\eeq
where we applied the reduction of the Nambu 3-bracket to a Poisson bracket

\beq
\{ f, g \}_{H_{2}} = \epsilon^{ijk} \partial^{j}f \partial^{k} g \partial^{i}H_{2} .
\eeq
By using the Fundamental Identity for $n=3$ we obtain

\beq
\{ \{ f , g \}_{H_{2}}, h \}_{H_{2}} + \{ \{ g , h \}_{H_{2}} , f \}_{H_{2}}  + 
\{ \{ h , f \}_{H_{2}} , g \}_{H_{2}}  = 0 ,
\eeq
the Jacobi identity for $ \{ , \}_{H_{2}} $. 

In order to get a consistent evolution for the coupled coordinates 
$ x^{i}, \ \ i=1,2,3 $ (eq. 3.26)  we must impose at $ t=0 $ 
the Poisson bracket algebras of the 
three coordinates

\beq
\{ x^{i}, x^{j} \}_{H_{2}} = \epsilon^{ijk} \partial^{k}H_{2} \ \ ; \ \ \ \ \  i,j,k=1,2,3 .
\eeq   
We observe that since $ H_{2} $ is a conserved quantity, the
 evolution eq.(3.27) preserves (3.30) in time. 

For $ H_{1}$ we choose a Hamiltonian describing the dynamics on the 2-dim. phase-space 
 $ H_{2}(x) = c $. Had we chosen $ H_{1}$ 
as the phase-space defining function then :

\beq
\frac{d x^{i}}{d t } = \{ x^{i}, H_{2} \}_{H_{1}}=- \{ x^{i}, H_{1} \}_{H_{2}} 
 \ \ ;\ \ \ \ \ \ \ \ \ \ i=1,2,3 .
\eeq
We get the time reversed evolution  if  we  impose  the  Poisson algebras

\beq
\{ x^{i} , x^{j} \}_{H_{1}} = \epsilon^{ijk} \partial^{k} H_{1} ,
\eeq
on the surface $ H_{1}(x) = c^{\prime} $ . 
The above interpretation of Nambu dynamics will provide the basic tool for the proposed 
quantization in section 6.

In the next section we shall connect Nambu dynamics 
in $ R^{3} $  with  flows $ SDiff(R^{3})$ and the NP 3-algebras with 
the infinite dimensional Lie algebras of  $ SDiff(R^{3})$.

%------------------------------------------------------------------------------------------

\section{ Volume Preserving Diffeomorphisms in the Clebsch-Monge Gauge \\
and Nambu Flows in $ R^{3}$}

%------------------------------------------------------------------------------------------

Since the famous paper by V.Arnold\cite{Arn} where he proved that the solution
of the Euler eqs. for perfect (incompressible and inviscid ) 
fluids \cite{EMar} are the geodesics of the infinite dimensional 
volume preserving diffeomorphism (VPD) group, 
there have been many developments.  In ref.\cite{MW} 
the symplectic structure discovered by Arnold was further studied and
a Hamiltonian formulation of the problem was proposed \cite{MRat}.

Here we will focus in the description of $ SDiff(R^{3})$, 
in a particular gauge, the Clebsch-Monge gauge, thus establishing 
the connection with Nambu Dynamics (flows) in $ R^{3} $.  
Our discussion easily extends to three dimensional  manifolds 
with a metric and  a smooth
Nambu tensor field.

Let $ {\cal A}=C^{\infty}(R^{3}) $ be the space of smooth functions on $ R^{3} $ and
$ {\cal G}=SDiff(R^{3}) $ be the set of smooth maps of $  R^{3} \mapsto R^{3} $ with the 
determinant of the Jacobian at each point of $ R^{3} $ equal to one, i.e.

\beq
J (f) (x) = det  [ \partial^{i} f_{j}(x) ] \ = \ 1  \ \ ; \ \ \ \ \ \ \ \ i,j=1,2,3 .
\eeq 
This set forms a group under composition of functions :

\beq
{\cal G} \times { \cal G} \ni (f , g ) \mapsto  f \circ g \in { \cal G } ,
\eeq
and the adjoint action of  is defined as :

\beq
Ad_{g} [f] = f \circ g^{-1} \ \ ; \ \ \ \ \ \ \ \forall f , g  \in { \cal G } .
\eeq
The  elements of the Lie algebra $ {\cal L(G)} $ are:

\beq
f^{i}(x) = x^{i} + v^{i}(x) \ \ ; \ \ \ \ \ \ \ \ \ \ \ \ i=1,2,3 \ ,
\eeq
with $ \partial^{i} v^{i} = 0 $.  We will impose conditions at infinity for 
$ v^{i}(x) $ : 

\beq
v^{i}(x) \stackrel{ |x| \rightarrow \infty}{\longrightarrow } 0 \ \ ; \ \ \ \ \ \ \ \ \ 
i=1,2,3 \ ,
\eeq
such that the total kinetic energy is finite (density constant) :

\beq
E = \frac{1}{2}\int \ \  d^{3}x \ \  v^{2}(x) \ \ \ < + \infty .
\eeq
For any infinitesimal element (4.4) we define the flow :

\beq
\frac{d x^{i}}{d t } = v^{i}(x) \ \ ; \ \ \ \ \ \ \ \ \ \ \ \ \ \ \ \ i=1,2,3 \ ,
\eeq
with initial conditions $ x^{i}_{o}= x^{i}(t=0) $ . 
Eq.(4.7) describes the motion of a particle which is immersed in a fluid 
of given stationary  velocity field  at the point $ x^{i}_{o} $, at $ t=0 $.

We can also define the fundamental representation of G on the space $ {\cal A}=C^{\infty}
(R^{3}) $ :

\beq
T_{g} (\alpha) = \alpha \circ g^{-1}  \ \ \ \ \ \ \ \ \alpha \in { \cal A } .
\eeq
By expanding for infinitesimal g :

\beq
g^{i}(x) = x^{i} + v^{i}(x) \ \ ; \ \ \ \ \ \ \ \ \ \ \ \ \ \ i=1,2,3 \ ,
\eeq
we get the action of generators

\beq
X(v) \alpha = - v^{i} \partial^{i} \alpha ,
\eeq
with a Lie algebra :

\beq
[ X(u) , X(v) ] = X(w) ,
\eeq
and composition law

\beq
w = ( u\cdot \partial ) v - (v \cdot \partial ) u = \partial \times ( u \times v ) .  
\eeq
Rewriting the flow eqs (4.7) via the use of the generators we get

\beq
\dot{x}^{i} = - X(v) \cdot x^{i}.
\eeq
We can integrate the equations of motion as :

\beq
x^{i}(t) = e^{- t \cdot X(v_{o})} \ \ x^{i}_{o} ,
\eeq
with $ v_{o} = v(x_{o})$ . 

After these basic preliminaries we introduce the Clebsch-Monge gauge 
\cite{MRat,Lam,Cleb,Poly1,Poly2,Mon}.
For every divergenceless vector field $ (v^{i}(x))_{i=1,2,3} \in R^{3} $, with boundary 
conditions of rel.(4.6)  we can find
a vector potential $ A^{i}(x) $ such that $ v^{i} = \epsilon^{ijk} \partial^{j} A^{k} $.
 Given $ A^{i}(x) $ Clebsch  and Monge  
introduced three
scalar potentials $ \alpha, \beta, \gamma  \in C^{\infty}(R^{3}) $ such that :

\beq
A^{i} = \partial^{i} \alpha + \beta \partial^{i} \gamma .
\eeq
So finally we get

\beq
v^{i}(x) = \epsilon^{ijk}\partial^{j}\beta \partial^{k}\gamma .
\eeq
The scalar function $ \alpha (x) $ becomes the gauge degree of freedom of $A^{i}(x)$.
From the last relation we see that the intersection of the surfaces $ \beta=\mbox{const.},
\gamma=\mbox{const.} $ define locally the flow lines. The existence of the scalar potentials
$ \beta, \gamma $ (Clebsch-Monge potentials)  is gurranteed locally if $ v^{i}(x) $ is
an analytic function in the region of a point say $ x^{i}=0,\ \ i=1,2,3 $. Then there
exists two integrals of motion of the flow equation :

\beq
\frac{d x^{i} } {v^{i}(x)} = dt \ \ ; \ \ \ \ \ \ \ \ \ \ i=1,2,3 \ ,
\eeq
f(x) , g(x) through which we can determine $ \beta $ and $ \gamma $.
The flows are characterized also by their vorticity

\beq
\omega^{i}(x) = \epsilon^{ijk} \partial^{j}v ^{k} \ \ ; \ \ \ \ \ \ \ \ \ \ i,j,k=1,2,3 .
\eeq
In case $ \omega^{i} = 0 $ the gradient flow $ v^{i} $ is :

\beq
v^{i}\ = \  - \partial^{i}\Phi \ \ ; \ \ \ \ \ \ \ \ \ \ i=1,2,3 ,
\eeq
where $ \Phi $ must be a harmonic function(Laplacian flow). 

In this case the surface $ \Phi = \mbox{const.} $ is orthogonal to the surfaces 
$ \beta =\mbox{const.} $ and $ \gamma=\mbox{const.} $ 
There are computer simulation studies of the flow eqs. 
for velocity fields general quadratic 
polynomials in the coordinates imposing zero radial motion on a 
sphere of radius R

\beq
\hat {n}\cdot v \mid _{|x|=R} = 0 .
\eeq
For various ranges of the polynomial coefficients one recovers  chaotic
flow as well as standard forms of flow modes \cite{Mof}.

Going back to rel (4.16) the generators of the flow , 
in terms of the Clebsch-Monge potentials,
become

\beq
X(\beta,\gamma ) \equiv  X ( \partial \beta \times \partial \gamma ) = - \epsilon^{ijk}
\partial^{j} \beta  \partial^{k} \gamma \partial^{i} ,
\eeq
The action of $ X(\beta,\gamma )$ on a smooth function $ \alpha \in C^{\infty}(R^{3}) $
is :

\beq
X ( \beta,\gamma ) \alpha = - \{ \alpha , \beta , \gamma \} ,
\eeq

the Nambu bracket of $ \alpha, \beta, \gamma $.
The flow eq(4.7) becomes

\beq
\dot {x}^{i} = \{ x^{i} , \beta , \gamma \} \ \ ; \ \ \ \ \ \ i=1,2,3 \ ,
\eeq
and so the Clebsch-Monge potentials of the flow are just the two Hamiltonians 
$ H_{1} = \beta ,\ \  H_{2} = \gamma $ of the Nambu dynamics .  We conclude that 
the flow equations  of incompressible fluids can be described by Nambu dynamics and vice versa.
 By considering
now the commutation relations (4.11-4.12)  in the Clebsch-Monge gauge
we obtain:

\beq
[ X(\beta_{1} , \gamma_{1} ) , X(\beta_{2} , \gamma_{2} ) ] = 
X( \{ \beta_{1} , \gamma_{1} , \beta_{2} \} , \gamma_{2} ) + 
X ( \beta_{2} , \{ \beta_{1} , \gamma_{1} , \gamma_{2} \} ) .
\eeq
Acting both sides of the CR(4.24),  on functions 
$ \alpha \in C^{\infty}(R^{3})$  we get the FI  :

\beq
\{ \beta_{1} , \gamma_{1} , \{ \beta_{2} , \gamma_{2} , \alpha \} \} -
\{ \beta_{1} , \gamma_{1} , \{ \beta_{2} , \gamma_{2} , \alpha \} \} =
\{ \{ \beta_{1} , \gamma_{1} , \beta_{2} \} , \gamma_{2} , \alpha \} + 
\{  \beta_{2} , \{ \beta_{1} , \gamma_{1} , \gamma_{2} \} , \alpha \} .
\eeq
We observe that all the information of the CR of $SDiff(R^{3})$ 
is contained in the NP 3-algebra for a basis of functions in $ R^{3} $ . 
Indeed if 
$ ( f_{\alpha} )_{\alpha \in S }$   
is a basis with index set S, then if we know the
structure constants of the 3-algebra , 
$ f^{\delta}_{\alpha \beta \gamma} $

\beq
\{ f_{\alpha} , f_{\beta} , f_{\gamma} \} = f^{\delta}_{\alpha \beta \gamma} f_{\delta}
\\ ; \ \ \ \ \ \ \ \ \alpha , \beta , \gamma , \delta \in S  ,
\eeq
then we can construct the Lie algebra structure constants for the  generators

\beq
X_{(\alpha , \beta )} = - \{ f_{\alpha} , f_{\beta} , \}  \ \ ; \ \ \ \ \  
\alpha , \beta \in S
\eeq
and commutation relations

\beq
[X_{(\alpha_{1} , \beta_{1} )} , X_{(\alpha_{2} , \beta_{2} )} ] = 
f_{\alpha_{1} \beta_{1}
\alpha_{2}}^{\gamma} X_{( \gamma , \beta_{2})} + f_{\alpha_{1} \beta_{1}
\beta_{2}}^{\gamma} X_{( \alpha_{2} , \gamma)}.
\eeq
Since later we shall need the case of linear or quadratic  Hamiltonians,
we give explicitly the construction of the corresponding NP 3-algebras. 
If both Hamiltonians are linear , i.e.   
$ H_{1}= a\cdot x, \ \ \ \  H_{2} = b\cdot x , \ \ \   a , b \in R^{3} $
then the flows

\beq
X( a , b ) = \epsilon^{ijk}\partial^{j} H_{1} \partial^{k} H_{2}
 \partial^{i} = ( a \times b)^{i} \partial^{i} \ \ ;
 \ \ \ \ \ \ \ \ \  i,j,k=1,2,3 \ ,
\eeq
represent translations along the direction $ a \times b $ (constant laminar flow).

If one is linear and the other is quadratic such as $ H_{1}= a  x  \ , \  
H_{2} = \frac{1}{2} x  B  x $ with $ \alpha\in R^{3} $ 
and B a real symmetric
$ 3 \times 3 $ matrix then :

\beq
X(\alpha, B ) = \epsilon^{ijk} a^{j} B^{kl} x^{l} \partial^{i} = ( A x )^{i} 
\partial^{i} \ \ ; \ \ \ \ \ i,j,k=1,2,3 \ ,
\eeq
with

\beq
A^{ij} = \epsilon^{ikl} a^{k} B^{lj} \ \ ; \ \ \ \ \ \ \ \ \ i,j,k,l=1,2,3 .
\eeq
It corresponds to a linear flow with an axis of symmetry $ a \in R^{3} $ . 
Finally if both Hamiltonians are quadratic : 
$ H_{1} = \frac{1}{2} x B x  \ , \   
H_{2} = \frac{1}{2} x C x $  with B,C real symmetric $ 3 \times 3 $ matrices 
(Quadratic flow) :

\beq
X(B,C) = \epsilon^{ijk} B^{jl} C^{km} x^{l} x^{m} \partial^{i} = A^{i}_{jk} x^{j} x^{k} 
\partial^{i} 
\eeq

\beq
A^{i}_{jk} = \epsilon^{ilm} B^{lj} C^{mk}.
\eeq
We denote  by  $ {\cal L_{C}(M) } \ , \ {\cal L_{L}(M)} \ , \  {\cal L_{Q}(M)} $ 
the constant, linear and quadratic flows respectively.  
It is easy to check that the commutator of 
elements of $ {\cal L_{Q}(M)} $ generate cubic flows. Hence only the sets    
$ {\cal L_{C}(M) } \ , \ {\cal L_{L}(M)}$  close by themselves under commutation. 
The associated commutation relations are :

\beq
[ X( a, b ) , X( c , d ) ] = 0 ,
\eeq

\beq
[X( a , b ) , X( c , B) ] = X( (c \times B) \cdot (a \times b ) ) ,
\eeq
and

\beq
[ X(a , A ) , X (b , B ) ] = X ( b , B \cdot ( a \times A ) ) - 
X ( a , A \cdot (b \times B )),
\eeq
where

\beq
( a \times A )^{ij} = \epsilon^{ilk} a^{l} A^{kj} \ \ ;
 \ \ \ \ \ \ \ \ \ \ i,j,k,l=1,2,3 .
\eeq

We proceed now to write down the CR  of  $SDiff(R^{3})$  
in the basis of plane waves from which we can generate the CR 
of any  other basis of $ C^{\infty}(R^{3})$.
We employ  linearity and Fourier transforms in order to 
consider the algebra of the exponential function 
$ e_{\alpha}= e^{i\alpha \cdot x}, \alpha \in R^{3} $ ( If $ \alpha \in Z^{3} $ 
we get the torus $T^{3}$ basis). 
The generators on this basis are :

\beq
X_{(\alpha , \beta )} = e_{\alpha + \beta} (\alpha \times \beta) \cdot \partial
\ \ ; \ \ \ \ \ \ \alpha , \beta \in R^{3}\ ,
\eeq
and we obtain:

\beq
\{ e_{\alpha} ,e_{\beta} , e_{\gamma} \} = - X_{(\alpha,\beta)\gamma} =
-i ( \alpha \times \beta) \cdot \gamma \ e_{\alpha + \beta + \gamma},
\eeq
so that

\beq
f_{\alpha\beta\gamma}^{\epsilon} = (-i)(\alpha \times \beta)\cdot \gamma 
\delta_{\epsilon-\alpha-\beta-\gamma} ,
\eeq

for $ \alpha, \beta, \gamma,\epsilon \in R^{3} $ . 
The Lie algebra of $SDiff(R^{3})$
on this basis becomes :

\beq
[ X_{(\alpha_{1} ,\beta_{1})} , X_{(\alpha_{1} ,\beta_{1})} ] = i ( \alpha_{1} \times 
\beta_{1} ) \cdot \alpha_{2}  X_{(\alpha_{1}+\beta_{1} + \alpha_{2} , \beta_{2})} +
i (\alpha_{1} \times \beta_{1} ) \cdot \beta_{2}  X_{(\alpha_{2} , \alpha_{1} + \beta_{1}
+ \beta_{2} )} .
\eeq

We close this section by the construction of the $SDiff(M_{3}) $ Lie algebra for a three
dimensional manifold $ M_{3} $ which can be embedded in $ R^{4}$  through a level set 
 function  $ h(x) = \mbox{const.}, \ \  \forall x \in  R^{4} $. 
For divergence free flows in $R^{4}$ ,

\beq
\partial^{a} v^{a} = 0 \ \ ; \ \ \ \ \ \ \  a=1,2,3,4 , 
\eeq
there exist three Clebsch-Monge potentials  $ \alpha, \beta, \gamma $ such that :

\beq
v^{a} = \epsilon^{a b c d } \partial^{b} \alpha \partial^{c} \beta \partial ^{d} \gamma
\ \ ; \ \ \ \ a,b,c,d=1,2,3,4 .
\eeq 
In order to define the incompressible flows on $M_{3}$ we consider the subset of flows on 
$ R^{4} $ with $ \gamma=h $. Then we set for the generators of the flow :

\beq
X_{h}(\alpha , \beta ) = \epsilon^{abcd} \partial^{b}\alpha \partial^{c}\beta 
\partial^{d} h \partial^{a} \ \ ; \ \ \ \ \ \ \ \ \ \ \ a,b,c,d=1,\cdots ,4 .
\eeq
For fixed h this defines a Lie subalgebra of $ SDiff(R^{4}) $ 
since $ X_{h}(\alpha,\beta) $
leaves invariant the manifold $ M_{3} \subset  R^{4} $ that is
the flow is parallel to $ M_{3}$ for points x of $ M_{3} $. 
The resulting subalgebra
is : 

\beq
[ X_{h}(\alpha_{1} , \beta_{1}) , X_{h}(\alpha_{2} , \beta_{2}) ] = 
X_{h} ( \{ \alpha_{1} , \beta_{1} , \alpha_{2} \}_{h} , \beta_{2} ) +
X_{h} ( \alpha_{2} , \{ \alpha_{1} , \beta_{1} , \beta_{2} \}_{h} ) ,
\eeq
with 

\beq
\{ \alpha , \beta , \gamma \}_{h} = \epsilon^{abcd} \partial^{b}\alpha 
\partial^{c} \beta \partial^{d} \gamma \partial^{a} h ,
\eeq
the induced 3-bracket from $ R^{4} $. Projecting on the
manifold $ {\cal M}_{3} $ we get the CR  of $ SDiff(M_{3}) $. 
Projection  in our present context implies the restriction of all functions 
$ \alpha_{1}, \alpha_{2}, \beta_{1}, \beta_{2} \in C^{\infty}(R^{4}) $ 
on the surface $ h(x)=\mbox{const.} $.
Since the generators $ X_{h}(\alpha, \beta) $ possess the Leibniz property with 
respect to $  \alpha_{1}, \alpha_{2}, \beta \in C^{\infty}(R^{4})$

\beq
X_{h}(\alpha_{1} , \alpha_{2} , \beta ) = 
\alpha_{1} X_{h}(\alpha_{2} , \beta ) + \alpha_{2}
X_{h}(\alpha_{1} , \beta ) ,
\eeq
it is enough to consider the CR only on the coordinate functions 
$ x^{a} \ \ \ , \ \ \ 
\alpha=1,2,3,4 $

\beq
 [ X_{h}( x^{a} , x^{b} ) , X_{h}(x^{c} , x^{d} ) ] = X_{h} ( \{ x^{a} , x^{b} , x^{c} \}_{h} ,
 x^{d} ) + X_{h} ( x^{c} , \{ x^{a} , x^{b} , x^{d} \}_{h},
 \eeq
where $ a,b,c,d =1,2,3,4 $ . 
Using the relation
 $ \{ x^{a}, x^{b}, x^{c} \} = \epsilon^{abcd} \partial^{d} h $  we  obtain :

\beq
[ X_{h}( x^{a}, x^{b} ), X_{h}( x^{c}, x^{d}) ] = \epsilon^{abcl} X_{h}(\partial^{l} h, 
x^{d} ) + \epsilon^{abdl} X_{h} (x^{c}, \partial^{l} h ) .
\eeq
If it is possible to solve parametrically the level-set eq. with smooth coordinate 
functions on $ M_{3} $ :

\beq
x^{a} = x^{a} ( \xi^{1} , \xi^{2} , \xi^{3} ) \ \ ; \ \ \ \ \ \ \ \ \ \ a=1,2,3,4 \ ,
\eeq
we obtain the Lie algebra of $ SDiff( {\cal M}_{3}) $ from eq.(4.41) for the 
coordinate function on $ {\cal M}_{3} $. For example, if h is a quadratic surface 
in $ R^{4} $ :

\beq
h = \frac{1}{2} x^{a} M^{a b} x^{b} \ \ ; \ \ \ \ \ \ \  \ a ,b = 1,2,3,4 \ ,
\eeq
where $ M^{ab} $ is a symmetric $ 4 \times 4 $ real matrix. For 
 $  a,b,c,d,l,k=1,2,3,4 $ we obtain :

\beq
[ X_{h}(x^{a} , x^{b} ) , X_{h}(x^{c} , x^{d}) ] = \epsilon^{abcl} M^{lk} 
X_{h} ( x^{k} , x^{d} ) +  \epsilon^{abdl} M^{lk} 
X_{h} ( x^{k} , x^{d} ) .
\eeq
If   M   is non-degenerate ( eigenvalues equal to  plus or minus one eigenvalues 
by diagonalizing and rescaling)
 we obtain the Lie algebra of the groups $ SO(p,q)  \ \ \ p+q=4 , \ \ p=1,2,3,4 $ 
for the 3-manifolds ${\cal M}^{p,q}_{3} $.

It becomes obvious from the previous observations  that Nambu dynamics 
can be represented as incompressible 
flows in a 3-d manifold $ {\cal M}_{3} $ and the NP 3-algebras are just
the Lie algebras of volume preserving diffeomorphisms 
of $ {\cal M}_{3} $. 
It is possible to restrict further the Nambu flows to the geodesics of 
  $SDiff({ \cal M}^{3}) $ so that the flows 
are solutions of the perfect fluid Euler equations.
In case we need higher dimensional embedding of 
$ {\cal M}_{3} $ to a $ R^{n} $ with 
$ n=2 \cdot 3 - 1 = 5 $ in general,   
we can extend our method to Nambu-Poisson 5-brackets and restrict with appropriate
level set functions $ h_{1} , h_{2}  $ to the manifold 
$ {\cal M}_{3} \subseteq  R^{5} $ .

%---------------------------------------------------------------------------------------

\section{Vortices in the Clebsch-Monge Gauge and their Topology in $ R^{3}$ }

%----------------------------------------------------------------------------------------

Flows contain  topological objects, the 3-d vortices 
and their interaction is governed by simple laws discovered by
H.von Helmholtz in 1858, Clebsch,Lord Kelvin, Poincare  and many others \cite{Saf,Arn1}.
 The topology of the vortex configurations 
in perfect barotropic fluids, is captured by the helicity \cite{WolMof} 

\beq
I = \frac{1}{(8\pi)^{2}} \int d^{3}x v^{i}(x)\omega^{i}(x) ,
\eeq
where the vorticity $\omega^{i} $ is defined by:

\beq
\omega^{i}(x) = \epsilon^{ijk}\partial^{j}v^{k} \ \ ; \ \ \ \ \ \ \ \ \ \ i,j,k=1,2,3 ,
\eeq
which is also divergenceless

\beq
\partial^{i} \omega^{i} = 0 .
\eeq
The helicity is a topological invariant of the flow and it is 
conserved in Euler inviscid flows.  
For applications in atmospheric fluid dynamics and condensed matter physics see,
for instance, 
\cite{NB} and \cite{PS} respectively.

It is possible to translate the divergenceless condition 
of the flow $ (v^{i}(x))_{i=1,2,3} $ 
to an algebraic constraint by introducing the nonlinear O(3) 
unit vector field $ (n^{i}(x))_{i=1,2,3}$ such that 
$ n^{i}n^{i} = 1 , (n \ \in \ S^{2}) $ \cite{Fad,KM}.
This is defined as follows (A is a dimensionful constant) :

\beq
\omega^{i} = A \epsilon^{ijk} \epsilon^{pqr} n^{p} \partial_{j} n^{q} \partial_{k}n^{r}
\ \ ; \ \ \ \ \ \ \ i,j,k,p,q,r=1,2,3 ,
\eeq
or vectorially

\beq
\omega^{i} = A \ \epsilon^{ijk} n \cdot (\partial_{j} n \times  \partial_{k} n) \ \ ;
 \ \ \ \ \ \ \ \ \ \ \ \ \ i,j,k=1,2,3 .
\eeq
It is easy to check that 
\beq
\partial^{i}\omega^{i} = det [ \partial^{i} n^{j}  ] = 0 .
\eeq
since $ (n^{i})_{i=1,2,3} $ are functionally dependent through $ n^{i}n^{i}=1 $.
The asymptotic condition (4.5) gurrantees that 

\beq
n^{i} \stackrel{|x|\rightarrow \infty }{\longrightarrow } n^{i}_{o} \in S^{2} ,
\eeq
the vector n approaches a constant vector $n_{o}$ as $ |x|$ goes to infinity. 
As a result we have a smooth map from $ R^{3}\sim S^{3} $ to $S^{2}$. The
Homotopy group of these maps is $ \Pi_{3}(S^{2})=Z $ and the integer 

%\beq
%\mbox {N} = \frac{1}{(8\pi )^{2}} \int d^{3}x v^{i}(x) \omega^{i}(x) ,
%\eeq
 Hopf invariant of the mapping 
$ n: S^{3}\rightarrow S^{2} $ 
is related to the helicity as : 

\beq
I \ \ = \ \  \mbox{N} A^{2} . 
\eeq
There is a nice geometrical interpretation of the Hopf integer number N in the flow 
picture given by  $ (n^{i}(x))_{i=1,2,3} $. 

Consider two fixed vectors $ n_{1}, n_{2} \in S^{2} $. For a particular field 
$ n(x)\in S^{2} $ let us follow the two vortex lines 
 $ n(x) = n_{i},\ \ i=1,2 $ for $ x \in  R^{3} $. 
Their linking number is precisely N. 
The vortex lines either  go  to infinity or must be closed.
If they are open and finite then $ \mbox{N} = 0$. 
In what follows we will discuss a particular parametrization of the incompressible flows
which results into a precise definition of the Nambu flows and brackets in the presence of
vorices. 
Given the topology of an incompressible flow it is possible to find locally a vector
potential $ ( A^{i})_{i=1,2,3} (x) $ :

\beq
\omega^{i} = \epsilon^{ijk} \partial^{j} A^{k}
\ \ ; \ \ \ \ \ \ \ \  i,j,k=1,2,3 .
\eeq
As discussed before  it is always possible to represent an arbitrary
vector field $A^{i}(\vec{x}) $ 
through three scalar potentials $ \alpha, \beta, \gamma $

\beq
A^{i} = \partial^{i}\nu + \mu \partial^{i}\lambda .
\eeq

The potential $\nu$ is the gauge freedom of rel.(5.11) 
and $\mu, \lambda$ are the Clebsch-Monge potentials, corresponding to the vorticity
$\omega^{i}$. 
%We have to solve their flow equations in order to determine them  :

%\beq
%\frac{d\beta}{dt} = \dot{x^{i}} \partial_{i}\beta = v^{i} \partial_{i}\beta .
%\eeq
%Similarly for $\gamma$ we have

%\beq
%\frac{d \gamma}{dt} = v^{i}\partial_{i}\gamma . 
%\eeq
We note here the important difference from the usual treatment of Euler flows \cite{MW}
where the Clebsch potential characterizes the vorticity 
$ \omega^{i}= \epsilon^{ijk}\partial^{j}\lambda \partial^{k}\mu $ 
rather than the velocity flow 
$v^{i} = \epsilon^{ijk} \partial^{j}\beta \partial^{k}\gamma $
which is our case of 
interest. 
In \cite{MRat} $ \lambda, \mu $ are canonical field variables for Euler flows.   

If there is a nontrivial topology in the flow (Hopf number $\neq 0 $)  we can determine 
$ \beta, \gamma$  by patching together the solution of the flow equation 
in different regions of $R^{3}$ .
The Clebsch-Monge  potential, $\beta$ or $\gamma$ are not single valued functions but rather
complicated non-local functions of $ \lambda,\mu $

The flow is expressed in terms of $\beta$ and $\gamma  $  and correspond to Nambu flows
 with Hamiltonians $ H_{1}=\beta, H_{2}= \gamma $:

\beq
v^{i} = \epsilon^{ijk} \partial^{j} \beta \partial^{k} \gamma \ \ ; \ \ \ \ \ \ \ \ \ 
i,j,k=1,2,3 .
\eeq
It can be shown that if $ \beta $ and $\gamma $ are single valued with the asymptotic
conditions for the velocity field (4.5) then the helicity  $ \mbox{N} =0 $.
The geometrical intersection
of the level surfaces 
$ \beta=c_{1}, \gamma=c_{2} \ \ \  \forall \ c_{1}, c_{2} \ \in \ R $, determines the
flow lines of the velocity field $\vec{v} $, implies that in the  case 
of a non-trivial topology the surfaces
$ \beta, \gamma $ must interwind each other. Hence it is natural that they are 
multivalued functions. 
This statement can be shown explicitly in terms of the unit vector 
$ (n^{i})_{i=1,2,3} $ 
introduced previously. 
We consider its polar angles $ \Theta(x), \Phi(x)  $ 

\beq
n = ( cos\Phi sin\Theta , sin\Phi cos\Theta , cos \Theta ) .    
\eeq
By calculating  $\omega^{i} $  we find: 

\beq
\omega^{i} = A \ \epsilon^{ijk} \ \partial^{j} cos\Theta \ \partial^{k}\Phi \ \ ;
 \ \ \ \ \ \ \ \ \ \ 
i,j,k=1,2,3 .
\eeq
We see that we can define ( set units A=1 ) :

\beq
\lambda = cos \Theta ,
\eeq
and 

\beq
\mu = \Phi ,
\eeq

We see the necessity of multivaluedness for $ \lambda ,\mu $ and therefore of $\beta$ and $ \gamma $ .  
The target manifold 
of $\beta$ and $\gamma$ at 
every space point, in general, may be a compact Riemann surface of arbitrary genus.
The symplectic structure of this space
leads to the non-uniqueness of $\beta$ and $\gamma$ in
representation of  the velocity field. Any area preserving transformation of $\beta $ and
$ \gamma $ on this surface 
leads to the same $v^{i}$. 
Representing  the vorticity $ \omega$ by Clebsch-Monge potentials $ \lambda, \mu $
the associated symplectic structure is precisely the
Arnold-Marden-Weinstein  structure on the space of functionals of 
vorticity \cite{MW,KM}.

%----------------------------------------------------------------------------------------

\section{ The Quantization of Nambu Dynamics in 3-D Phase Space }

%-----------------------------------------------------------------------------------------

In section 2 we stressed the importance of the properties of the Nambu 3-bracket, 
such as 
a) Leibniz , and b) the Fundamental Identity(FI) 
for the consistency of the classical evolution eqs. of Nambu mechanics(NM) 
in 3-d  manifolds.  
Focusing our discussion on $R^{3}$ (although it is
easily generalizable to 3-manifolds embeddable in $R^{4}$) 
our interpretation of section 3, is that we choose among the two Hamiltonians 
$H_{1}$ or $ H_{2}$  
\footnote{In this work we restrict ourselves to the space of polynomials 
of coordinates for the Hamiltonians $H_{1}, H_{2}$.} 
the one which  defines the 2-d phase space geometry embedded in $R^{3}$, say 
$ H_{2}(x)=C $. 
For various initial conditions we obtain a foliation of 
$R^{3}$ into two dim. phase spaces all possessing the same Poisson algebra
of coordinates at $ t=0 $ 

\beq
\{ X^{i} , X^{j} \}_{H_{2}} \ = \ \epsilon^{ijk} \partial^{k}H_{2} \ \ ;
 \ \ \ \ \ \ \ \ \ \ 
\ \ \ i,j,k=1,2,3 .
\eeq
The second Hamiltonian $ H_{1}$ defines the dynamics of the 
motion on the $H_{2}$ phase-space :

\beq
\dot{X}^{i} \ = \ \{ X^{i} , H_{1} \}_{H_{2}} \ \ ; \ \ \ \ \ \ \ \ \ \ \ \ \ \  i=1,2,3 .
\eeq
Since $ H_{2} $ is conserved, for all later times the phase space 
coordinates satisfy the same algebra : 

\beq
\{ X^{i}(t , x_{0}) , X^{j}(t ,x_{0}) \} \ = \ \epsilon^{ijk} \partial^{k}H_{2}.
\eeq
We propose an almost obvious quantization rule for NM as follows. 

We, firstly, define an associative quantization of the algebra (6.1) promoting the
phase space coordinates $X^{i}$ at $t=0$ 
to hermitian operators with commutation relations (CR) :

\beq
[ X^{i} , X^{j} ] \ = \ X^{i}X^{j} - X^{j}X^{i} =  
\imath \hbar\epsilon^{ijk}P^{k}(x) \ \ ; \ \ \ \ \ \ \ \ \ \ \ \ \ i,j,k=1,2,3 ,
\eeq
having as a classical limit 

\beq
\mbox {lim} \ \frac{1}{i\hbar} [ X^{i} , X^{j} ] \ \stackrel{\hbar \rightarrow 0}{=} \ 
\{ X^{i} , X^{j} \}_{H_{2}} , 
\eeq 
or

\beq
\mbox{lim} \ P^{k}(x) \ \stackrel{\hbar \rightarrow 0}{=} \ \partial^{k} H_{2}(x).
\eeq

If $H_{2}$ is a quadratic function of the canonical phase space coordinates there is
no ordering problem (linear Lie-algebras). For $H_{2}$ cubic or higher (non-linear
Lie algebras) there is no unique way to quantize. Nevertheless the polynomials
$ P^{k}(x) \ \ , \ \ k=1,2,3 $ must obey the following constraints :

a) They must be  hermitian operators (e.g. by Weyl ordering of $\partial^{k}H_{2} $)

b) They must satisfy the Diamond Lemma \cite{Newm}. 
The algebra (6.4) must have a Universal 
envelopping algebra ${\cal U}$, for which any monomials of 
$ X^{i}, (X^{i})^{m_{1}} (X^{j})^{m_{2}}(X^{k})^{m_{3}} $ 
can be brought using the polynomial commutation relations 
to a prechosen order such as for example
$(X^{1})^{n_{1}}(X^{2})^{n_{2}}(X^{3})^{n_{3}}$.

This property is necessary for the existence of a basis  of ordered monomials 
of  ${\cal U}$
as well as for comparisons of LHS and RHS respectively of various identities. 
This is analogous to the Poincare-Birkoff theorem, for linear Lie algebras.

c) They must obey the Jacobi identity 

\beq
[ X^{1} , P^{1} ] + [ X^{2} , P^{2} ] + [ X^{3} , P^{3} ] = 0 ,
\eeq
and finally

d) There  must exist a Casimir for the algebra (6.4) $ H_{2}(\hbar ) $ 

\beq
[ X^{i} , H_{2}( \hbar ) ] = 0 ,
\eeq
such that the Classical limit exists and moreover 

\beq
\mbox{ lim} H_{2}(\hbar ) \stackrel{\hbar \rightarrow 0}{=} H_{2},
\eeq
where $H_{2}$ is the classical Casimir.

Non-linear Lie algebras have been discussed as deformations of linear Lie algebras
(Quantum Groups, W-algebras , polynomial Lie algebras)\cite{CP,QGr,Walg}.

The cohomological obstruction for $\star$-quantization of polynomial Poisson algebras
has been studied in ref.\cite{Now}. 
Recently in ref.\cite{ABHS} a framework has been proposed for matrix deformations, 
corresponding to non -linear Poisson algebras for compact surfaces in $R^{3}$ of any genus.  
Explicit constructions, as far as we know, have been
given only for deformed spheres $g=0$ and  tori $g=1$.  

Once we have quantized the algebra of phase space coordinates at 
$t=0$ with Casimir $H_{2}(\hbar )$  
we proceed to introduce the following quantum Nambu-Heisenberg
eqs. : 

\beq
\imath \hbar \frac{d X^{i}}{d t} \ = \ [ X^{i} , H_{1} ]_{H_{2}(\hbar )} 
\ \ ; \ \ \ \ \ \ \ \ \ i=1,2,3 ,
\eeq
where the commutator on the RHS  has to be evaluated with the quantum algebra (6.4).
We observe that since the commutator respects the Leibniz property for any observable 
 F  which is not explicitly dependent on time we obtain the quantum Liouville eqn.:

\beq
\imath \hbar \frac{d F(X)}{d t} = [ F , H_{1} ]_{H_{2}(\hbar )}.
\eeq
In particular $ H_{1} $ and $ H_{2}(\hbar ) $ 
are conserved and thus $ X^{i} $ satisfy the
same algebra for all times :

\beq
[ X^{i}(t,x_{0}) , X^{j}(t , x_{0}) ] = \imath \hbar \epsilon^{ijk} P^{k}(X) \ \ ;
 \ \ \ \ \ \ \ \ \ i,j,k=1,2,3 .
\eeq

We can formally solve eq.(6.10) by using the adjoint operator $\mbox{ad}_{X}$

\beq
\mbox{ad}_{X}[Y] = [Y,X] ,
\eeq

\beq
F(X) \ = \ e^{- \frac{\imath} { \hbar } t ad_{H_{1}} } \ F(X_{0})  \ = \ 
e^{- \frac{\imath }{\hbar }t H_{1}} F(X_{0}) e^{ \frac{\imath }{\hbar }t H_{1}}.
\eeq

We end this section by providing three illustrative examples for our construction.
1)  An electric charge in a homogeneous magnetic field. 
The classical phase space is defined by the $ H_{2} $ function :

\beq
H_{2} \ = \ \frac{e}{m^2 c} \vec{v} \cdot \vec{B},
\eeq
and so the Nambu-Poisson algebra of the phase-space coordinates $ v^{i} $ is 
according to rel.(6.1),

\beq
\{ v^{i} , v^{j} \} =  \frac{e}{m^2 c} \epsilon^{ijk}B^{k} \ \ ; 
\ \ \ \ \ \ \ \ i,j,k=1,2,3 .
\eeq
The phase space is a plane transverse 
to B embedded in $ R^{3} $ . The dynamics is defined
through : 

\beq
H_{1} \ = \ \frac{1}{2} m v^{2},
\eeq
and the Nambu eqs: 

\beq
\dot{v}^{i} \ = \ \frac{e}{mc} \epsilon^{ijk} v^{j} B^{k},
\eeq
produce the correct physical eqs. of motion. 
For the quantum case we have the following two Hamiltonian operators : 

\beq
\hat {H}_{2} \ = \ \frac{e}{m^2 c} \hat{v} \cdot B,
\eeq
and 

\beq
\hat{H}_{1} \ = \ \frac{1}{2} m \hat{v}^{2}.
\eeq
For the algebra of coordinates we get a Heisenberg Lie algebra : 

\beq
[ \hat{v}^{i} , \hat{v}^{j} ] \ = \ \imath \hbar \frac{e}{m^2 c} \epsilon^{ijk} B^{k}
\ \ ; \ \ \ \ \ \ \ \ \ i,j,k=1,2,3.
\eeq

$ \hat{H}_{2} $ is the Casimir of the Heisenberg algebra which defines the 
quantum plane foliating $ R^{3} $.

The Nambu-Heisenberg eqs. of motion are :
\beq
\frac{d \hat{v}^{i}}{d t} \ = \ \frac{e}{mc} \epsilon^{ijk} \hat{v}^{j}B^{k} =
- \frac{\imath }{\hbar } [ \hat{v}^{i} , \hat{H}_{1} ]_{\hat{H}_{2}}.
\eeq  
These are the standard QM eqs. for the Landau problem \cite{Lan}.

2)  The Euler Top \cite{Nam}
At the classical level we choose 

\beq
H_{2} \ = \ \frac{1}{2} \ l^{i} \ l^{i}. 
\eeq
The corresponding phase space is $ S^{2} $ 
which provides a spherical foliation of $ R^{3} $ with varying
radius $ \sqrt{2H_{2}} $ for various initial conditions $l^{i}_{0} $ 
with Poisson
algebra $ SO(3) $ 

\beq
\{ l^{i} , l^{j} \} = \epsilon^{ijk} \ l^{k} \ \ ; \ \ \ \ \ \ \ \ \ \ i,j,k=1,2,3 .
\eeq
The second Hamiltonian is the conserved energy

\beq
H_{1} \ = \ \frac{1}{2} \ ( \frac{l_{1}^{2} }{I_{1}} + \frac{l_{2}^{2} }{I_{2}} +
\frac{l_{3}^{2} }{I_{3}} ).
\eeq
The classical eqs. of motion are 
$ \dot{l}^{i} = \epsilon^{ijk} \ \partial^{j}H_{1} \ \partial^{k}H_{2}$ or

\beqa
\dot{l}^{1} &=& ( \frac{1}{I_{2}} - \frac{1}{I_{3}} ) \ l_{2} \ l_{3}  \nonumber \\
\dot{l}^{2} &=& ( \frac{1}{I_{3}} - \frac{1}{I_{1}} ) \ l_{3} \ l_{1}            \\
\dot{l}^{3} &=& ( \frac{1}{I_{1}} - \frac{1}{I_{2}} ) \ l_{1} \ l_{2}  \nonumber .
\eeqa
In the quantum case 

\beq
\hat{H}_{2} \ = \ \frac{1}{2} \ \hat{l}^{i} \ \hat{l}^{i} \ \ ;
 \ \ \ \ \ \ \ \ i=1,2,3 .
\eeq

The phase-space Lie algebra is linear ($ SO(3) $)

\beq
[ \hat{l}^{i} , \hat{l}^{j} ] = \imath \hbar \ \epsilon^{ijk} \ \hat{l}^{k} \\ ; 
\ \ \ \ \ \ \ \ \ \ \ \  i,j,k=1,2,3 .
\eeq
The Energy operator is $ H_{1} $

\beq
\hat{H}_{1} \ = \ \frac{1}{2} ( \frac{\hat{l}_{1}^{2} }{I_{1}} + \frac{\hat{l}_{2}^{2} }{I_{2}} +
\frac{\hat{l}_{3}^{2} }{I_{3}} ).
\eeq

The quantum Nambu-Heisenberg eqs. of motion are: 

\beq
\imath \hbar \frac{d \hat{l}^{i}}{d t} = [ \hat{l}^{i} , H_{1} ]_{H_{2}} \ \ ; 
\ \ \ \ \ \ \ \ \ \ i=1,2,3 ,
\eeq

or component wise 

\beqa
\frac{d \hat{l}^{1}}{d t} &=& \frac{1}{2} (\frac{1}{I_{2}} - \frac{1}{I_{3}} ) 
(\hat{l}_{2}\hat{l}_{3} + \hat{l}_{3}\hat{l}_{2}) \nonumber \\
\frac{d \hat{l}^{2}}{d t}  &=& \frac{1}{2} ( \frac{1}{I_{3}} - \frac{1}{I_{1}} ) 
(\hat{l}_{3}\hat{l}_{1} + \hat{l}_{1}\hat{l}_{3})           \\
\frac{d \hat{l}^{3}}{ d t}  &=& \frac{1}{2} (\frac{1}{I_{1}} - \frac{1}{I_{2}} ) 
(\hat{l}_{1}\hat{l}_{2} + \hat{l}_{2}\hat{l}_{1} ) \nonumber .
\eeqa

These are the correct eqs. of motion for the quantum top\cite{Lan}. It is known
that the prescription of quantization by Nambu\cite{Nam}
for the quantum triple product fails by a multiplicative factor on the RHS of eq.(6.31)
which is the value of the $ SO(3)$ Casimir

3)  Single Spin Magnetic Field Interaction
This example is similar in spirit to the first one describing the motion of a quantum
particle of magnetic moment $ \mu $ and quantum spin  s 

\beq
M^{i} = \mu \hat{S}^{i} \ \ ; \ \ \ \ \ i=1,2,3 \ ,
\eeq
with Hamiltonians $ H_{2} $ and $ H_{1} $ 

\beqa
\hat{H}_{2} &=& \frac{1}{2} \hat{S}^{i} \hat{S}^{i}   \nonumber \\
\hat{H}_{1} &=& -\mu B^{i} \hat{S}^{i} \ \ ; \ \ \ \ i=1,2,3 .
\eeqa

The phase space algebra is $ SU(2)$ 

\beq
[ \hat{S}^{i} , \hat{S}^{j} ] \ = \ \imath \hbar  \ \epsilon^{ijk} \ \hat{S}^{k}
\eeq

with the corresponding eqs. of motion being : 

\beq
\imath \hbar  \frac{d \hat{S}^{i}}{ d t} \ = \ [ \hat{S}^{i} , \hat{H}_{1} ]_{\hat {H}_{2}} ,
\eeq

or equivalently 

\beq
\frac{d \hat{S}^{i}}{d t} = - \mu \epsilon^{ijk} B^{j} \hat{S}^{k} \ \ ; 
\ \ \ \ \ \ \ \ \ \ i,j,k=1,2,3 \ ,
\eeq

which again are the expected ones. 

We note that in these three examples and for general quadratic or 
linear polynomial Hamiltonians $ \hat{H}_{1} , \hat{H}_{2} $ it is easy to check that 

\beq
[ \hat{X}^{i} , \hat{H}_{1} ]_{\hat{H}_{2}} \ = \ - 
[ \hat{X}^{i} , \hat{H}_{2} ]_{\hat{H}_{1}} .
\eeq

Note that the exchange symmetry $ \hat{H}_{1} \leftrightarrow \hat{H}_{2} $ 
between the two Hamiltonians is equivalent
to time reversal symmetry $ t \rightarrow -t $ . More generally this duality symmetry
is valid for any element $ g \in SL(2,R) $  

\beq
g = \left( \begin{array}{cc} \alpha \ \ \ \beta \\ \gamma \ \ \ \delta 
\end{array}\right) \ \ ; \ \ \ \ \ \ \ \ \ \ \mbox{det g }=1 , 
\ \ \ \ \ \ \alpha , \beta ,\gamma , \delta \ \in \ R \ ,
\eeq
  
which produces the transformation 

\beq
(\hat{H}_{1} , \hat{H}_{2} ) \rightarrow  \ \ \ \ \ \ 
( \hat{H}_{1}^{\prime } , \hat{H}_{2}^{\prime} ) \ = \ (\hat{H}_{1} , 
\hat{H}_{2} ) \cdot g .
\eeq

For general quadratic Hamiltonians it leaves invariant the equations of motion

\beq
\imath \hbar \frac{d \hat{X}^{i} }{d t} =  [ \hat{X}^{i} , \hat{H}_{1} ]_{\hat{H}_{2}}
\ \ ; \ \ \ \ \ \ \ i=1,2,3 .
\eeq
The general setting we have developed here is appropriate to the quantization of
classical flow eqs. for perfect fluids (see discussion in section 5). 
For many years this is a very active field starting 
with Landau (1941)\cite{Land,Fey,Vol}. He formulated
Quantum Hydrodynamics in the Eulerian framework by quantizing the density 
$ \rho $ and
the current $ J^{i}, \ \ \  i=1,2,3 $ starting from basic 
commutation relations of flow coordinates for the constituent particles 
( Lagrangian formulation). The physical
phenomenon at hand was superfluidity and more specifically He4\cite{Khal}. 
In the last two decades, there has been an intense interest  for quantum fluids 
( BEC )\cite{Legg} 
and strongly
correlated electron systems(quantum Hall effect and 
high temperature supercontuctivity) \cite{QHall}. 
On the other hand for studies related to
non-commutative or fuzzy fluids see ref.\cite{Poly2,Raj}. In addition
 very recently there has been a very fruitful connection of $AdS_{5}$ 
black hole geometry with the quark-gluon fluid
 thermodynamics on the boundary \cite{SS}.

Having established the precise physical setting of our proposal 
we proceed to discuss in the next section the quantization of 
the Nambu-Poisson 3-algebras (3-brackets). According
to our approach it must be consistent with the quantum 
Nambu-Heisenberg equations of motion. Few of the works 
in the literature have made a consistent connection of the quantization
of the Nambu  3-bracket with Quantum Nambu Dynamics. 

%-----------------------------------------------------------------------------------------

\section{Nambu-Lie 3-Algebras and the Quantization of the 3-Bracket}

%-----------------------------------------------------------------------------------------

Nambu-Lie 3-algebras have been discussed  in the very
past ref. \cite{ Nam,Fil,Tak,CZ2,DTak},
and more recently as metric linear 3-algebras \cite{papado,L3alg}.
 They are defined as algebras with a finite set of generators 
$T^{a}, \ \ a=1,2,\cdots , n $
and a 3-commutator with the following properties:

1) Antisymmetry 

\beq
[ t^{\sigma(a)} , t^{\sigma(b)} , t^{\sigma(c)} ] \ = 
\ (-1)^{\sigma} \ \ \  [ t^{a}, t^{b},t^{c} ]  \ \ ;  \ \ \ \ \ \ \ 
 a,b,c=1, \cdots, n ,
\eeq

for every permutation of three objects $ \sigma  \ \ \in  \ \  S_{3} $ 

2) Linearity

\beq
[ \lambda_{a}t^{a} , t^{b} , t^{c} ] \ = \ \lambda_{a} [ t^{a} , t^{b} , t^{c} ] 
\ \ ; \ \ \ \ \ \  \lambda_{a} \in {\cal C }, \ \ \ \ \  a,b,c = 1,\cdots , n ,
\eeq

3) Fundamental Identity(FI)

\beqa
[ [ t^{a} , t^{b} ,t^{c} ] , t^{d} , t^{e} ] \ &=& \ 
[ [ t^{a} , t^{d} ,t^{e} ] , t^{b} , t^{c} ] + 
[  t^{a} , [ t^{b} ,t^{d} ] , t^{e} ] , t^{c} ] + \\ & &
[  t^{a} , t^{b} , [ t^{c}  , t^{d} , t^{e} ] ] \ \ ; \ \ \ \ \ \ \ 
\forall a,b,c,d,e=1,2, \cdots, n .
\eeqa

The last property can be expressed in a different way . 
If we define the adjoint action operator :

\beq
L_{a,b} \ \equiv \ [ t^{a} , t^{b} , ] \ \ ; \ \ \ \ \ \ \ \ \ \forall \ 
a,b=1,\cdots ,n .
\eeq
It acts like a derivation on  the 3-commutator : 

\beq
L_{d,e} [ t^{a} , t^{b} , t^{c} ] \ = \ [ L_{d , e} t^{a} , t^{b} , t^{c} ] +
[ t^{a} , L_{d,e} t^{b} , t^{c} ] + [ t^{a} , t^{b} , L_{d,e} t^{c} ].
\eeq 
It generalizes the usual action of the adjoint operation of a Lie algebra
or equivalently it is an extension of the Jacobi identity.
A question of consistency is in order, when the
Leibniz property is imposed in addition to the previous ones :

4) Leibniz

\beq
[ t^{a} , t^{b} , t^{c} , t^{d} ] = t^{a} [ t^{b} , t^{c} , t^{d} ] +
[ t^{a} , t^{c} , t^{d} ] t^{b} .
\eeq
It is possible to construct 3-algebras which do not satisfy the FI but they do 
instead satisfy the Leibniz property (Leibniz 3-algebras) \cite{DTak}.
The latter is necessary in order  to extract 
from the 3-commutator the generators of the algebra the 3-commutator 
of the polynomials in the generator, 
in other words the full structure of the enveloping algebra $ {\cal U}$.

The final property is 

5) The Closure relation 

\beq
[ t^{a} , t^{b} , t^{c} ] = i f^{abc}_{d} t^{d} \ \ ; 
\ \ \ \ \ \ \ \ \ a,b,c,d=1,\cdots , n .
\eeq

To write down a Lagrangian one also needs an inner product trace form which raises and
lowers indices on the algebra $ \mbox{Tr} (t^{a} t^{b}) = h^{ab} $ . 
These algebras are called
"Metric Lie 3-algebras". 

We name the algebras which satisfy properties 1)-5), as "Linear Nambu-Lie 3-algebras" 
 in order to distinquish their structure from more general Non-linear Nambu-
Lie 3-algebras 

\beq
[ t^{a} , t^{b} , t^{c} ] = i f^{abc}_{d} P^{d}(t) , 
\eeq

where $ P^{d}(t) \ , \ d=1,\cdots, n $ are polynomials in the 
generators $ t^{a} $ . The FI
imposes constraints on the $ f^{abc}_{d} $ and in the more general case on the Polynomials 
$ P^{d} $ . 

In the BL theory \cite{BL,Gust} the Leibniz property is ignored because 
it is not necessary for the consistency of the theory. 
The Leibniz property itself assumes the existence of a
product between the generators which can be associative or 
non-associative although some properties are in directly assumed 
at the level of traces. In the literature there are proposals for the 
3-commutator which start directly from a triple product between
the generators.  For cubic matrix algebras \cite{Mats,Kaw} as well as for non-associative 
3-algebras  one starts off from the
associator 

\beq
< t^{a} , t^{b} , t^{c} > = t^{a}(t^{b}t^{c}) - (t^{a}t^{b}) t^{c} ,
\eeq

The 3-commutator bracket is then defined to be:

\beq
[ t^{a} , t^{b} , t^{c} ] = \sum_{\sigma \in S^{3}} \ \ (-1)^{\sigma} 
< t^{\sigma(a)} , t^{\sigma(b)} , t^{\sigma(c)} > . 
\eeq 
The well known non-associative algebra of octonions (7-imaginary units) 
$ e_{i}, \ \ i=1,\cdots , n=7  $
satisfy \cite{octon,CS}

\beqa
e_{i}e_{j} & = & -\delta_{ij} + \Psi_{ijk} e_{k} \ \ \ \ \ \ \ 
i,j,k=1,\cdots , 7 \nonumber \\ 
e_{0}e_{i} & = & e_{i}e_{0} \ \ \ \ \ \  i=1, \cdots ,7   \\
e_{0}^{2} &=& 1 .
\eeqa
The associator is given by 

\beq
< e_{i} , e_{j} , e_{k} > = e_{i}(e_{j}e_{k}) - ( e_{i}e_{j} ) e_{k} = 
\varphi_{ijkl} e_{l} \ \ \ \ \ \ \ \ \ \ i,j,k,l=1,\cdots ,7 ,
\eeq
where $ \Psi_{ijk} $ is the completely antisymmetric tensor of octonionic 
multiplication table with values 
1 for  [(123), (246),(435),(367),(651),(572),(714)]  
and zero otherwise.The dual tensor $\varphi_{ijkl}$ is
defined as

\beq
\varphi_{ijkl} = \epsilon_{ijklmnp} \Psi_{mnp} \ \ ; 
\ \ \ \ \ \ \ \ i,j,k,l,m,n,p=1,\cdots ,7 .
\eeq
It is  completely antisymmetric  with values 
1 for $ (1245),(2671),(3526), \\
(4273),(5764),(6431),(7531)$   and zero otherwise. 
The seven octonionic units form a linear 3-algebra which  is given by

\beq
[ e_{i} , e_{j} , e_{k} ] = 7 \ \ \varphi_{ijkl} e_{l} \\ ; 
\ \ \ \ i,j,k,l= 1 , \cdots ,7 ,
\eeq

but it does not satisfy the FI and Leibniz properties.
We would like to notice here the relation of octonions with the quantum mechanical
self-dual membranes (instantons), in the light-cone gauge, embedded in 
7 dimensions \cite{Fleo,Fair}.
For associative linear NL 3-algebras the triple commutator is 

\beq
[ t^{a} , t^{b} , t^{c}] \ = \ \sum_{\sigma \in S^{3}} (-1)^{\sigma} 
t^{\sigma(a)} , t^{\sigma(b)} , t^{\sigma(c)} .
\eeq
In order to define the triple commutator, one could also choose an element 
$ \Gamma , \Gamma^{2}=I $ such that 

\beq
[ \Gamma , t^{a} ]_{+} = 0 .
\eeq
The 3-commutator is then defined through the 4-commutator \cite{CZ,Jab2}

\beq
[ X^{a} , X^{b} , X^{c} , X^{d} ] =  \sum_{\sigma \in S^{4}} (-1)^{\sigma} 
X^{\sigma(a)} X^{\sigma(b)} X^{\sigma (c)} X^{\sigma(d)} ,
\eeq
as:

\beq
[t^{a} , t^{b} , t^{c} ] \equiv [ t^{a} , t^{b} , t^{c} , \Gamma ].
\eeq
It has been proved that the closure relation (7.8) for positive definite 
metric 3-algebras 
has solutions only for $n=4$, the $ A_{4}$  algebra  or  direct sums   with
abelian triple algebras\cite{papado}. 

The $ A_{4} $ algebra has as generators\cite{BL}

\beq
t^{a} = \gamma^{a} \ \ ; \ \ \ \ \ \ \ \ \ \ \ \ \ \ a=1,2,3,4 ,
\eeq
and $ \Gamma = \gamma^{5} $, (two SU(2) algebras of positive and negative chirality):  

\beq
[ t^{a} , t^{b} , t^{c} ] = i \epsilon^{abcd} t^{d} \ \ ;  \ \ \ \ \ \ \ \ 
\ \ a,b,c,d=1,2,3,4 .
\eeq
 In general the definitions of the triple 
commutator (7.10, 7.11, 7.17, 7.19 , 7.20) do not
satisfy the FI and Leibniz properties. 

As has been emphasized in the previous sections, our approach is to consider 
Nambu-Lie 3-algebras which allow for the consistent quantization of 
Nambu classical dynamics in 3-d phase-space manifolds $ {\cal M}_{3} $.
 This, in turn means (see sect. 3-4), that we should quantize consistently 
the Lie algebras of volume preserving diffeomorphisms in the Clebsch-Monge(CM) gauge. 
One way would be to quantize the CM potentials
as we do in quantum field theory, 
by using familiar symplectic structures  \cite{MW,Arn}.
A second way would be, to construct  topological $\sigma$-models defining  
 the $*$ deformation of the Poisson
algebra of smooth functions on $ {\cal M}_{3} $ \cite{Cat}.

Our approach is to consider Matrix deformations of the algebras of 
coordinates for every surface defined by a level set Morse function, which
is the Casimir of the corresponding Poisson algebra( see section three).
 In accord with our philosophy of sect.5 we have to be consistent with the 
Nambu-Heisenberg equations of motion. If we choose the two
Hamiltonians $ \hat{H}_{1} , \hat{H}_{2} $ then the time evolution equations are

\beq
\imath \hbar \frac{d \hat{X}^{i}}{d t} = \ [ \hat{X}^{i} , \hat{H}_{1} ]_{\hat{H}_{2}}.
\eeq
We define the Nambu quantum 3-bracket as the 3-commutator 

\beq
[ \hat{X}^{i} , \hat{H}_{1} , \hat{H}_{2} ] \ = \ [ \hat{X}^{i} , 
\hat{H}_{1} ]_{\hat{H}_{2}} .
\eeq
Any polynomial Hermitian operator observable $ \hat{F}(\hat{x}) $ 
satisfies the Quantum Liouville time evolution equation generically due to 
our  ansatz

\beq
\imath \hbar \frac{d \hat{F}}{d t} = [ \hat{F} , \hat{H}_{1} ]_{\hat{H}_{2}}.
\eeq 
It also  follows from (7.22) that more generally we have

\beq
[ \hat{F} , \hat{H}_{1} , \hat{H}_{2} ] = [ \hat{F} , \hat{H}_{1} ]_{\hat{H}_{2}}.
\eeq
The triple commutator just defined, if used for any three Hermitian operators
F, G, H (we omit hats from now on) :

\beq
[F , G , H ] = [ F , G ]_{H} ,
\eeq

obeys as before the following properties: a) Linearity b) Antisymmetry 
c)Leibniz in the first two arguments. If the additional requirement is imposed, namely that

\beq
[ F , G ]_{H} \ = \ - [ F , H ]_{G} ,
\eeq
all of the above properties get satisfied as well in all three arguments. 
By fixing 
the phase space 
to be $R^{3}$ we will examine rel.(7.28) for the case that the three operators 
F,G,H are linear
or quadratic in the coordinates $x^{i}$. 

1) Linear case

\beq
F= a^{i} x^{i} , \ \ \ \ \ G =b^{j} x^{j} , \ \ \ \ \ H= c^{k} x^{k} \ \ ; 
\ \ \ \ \ \ a,b,c \in R^{3}, \ \ \ \ \ i,j,k=1,2,3.
\eeq
According to our definitions the algebra of coordinates is :

\beq
[ x^{i} , x^{j} ]_{H=c^{k}x^{k}} \ = \ \imath \hbar  \epsilon^{ijk}c^{k}.
\eeq
This is the non-commutative 3-torus $ T^{3}_{c} $ \cite{Rief}. 
Since the Casimir H defines a quantum plane 
( the usual quantum mechanical phase-space ) for every value $ \lambda $ 
of an irrep: 
$ \lambda \in R $ 

\beq
H = c^{k} x^{k} = \lambda \cdot I .
\eeq
The non-commutative 3-torus is foliated by the $\lambda$-planes(2-tori)\cite{AFN1}. 
We find for the commutator $ [ F, G ]_{H} $ 

\beq
[ a^{i} x^{i} , b^{j} x^{j} ]_{c^{k}x^{k}} \ = \ \imath \hbar a \cdot ( b \times c ).
\eeq

Hence rel.(7.28) holds true, as the RHS of (7.32) is antisymmetric in 
$ b \leftrightarrow c $ 

2) F,G  Linear  ,  H Quadratic

\beq
H \ = \ \frac{1}{2} x^{k} M^{kl} x^{l} \ \ ; \ \ \ \ \ \ \ \ k,l=1,2,3 ,
\eeq

where M is a real symmetric matrix. The algebra of coordinates is a 
3-generator linear Lie algebra. Depending on the eigenvalues of M 
we obtain all cases ( SU(2) , SU(1,1), etc.). 

\beq
[ x^{i} , x^{j} ]_{H} \ = \ \imath \hbar  \epsilon^{ijk} M^{kl} x^{l}.
\eeq
Foliating $ R^{3} $ by fuzzy quadratic surfaces the LHS of rel.(7.28) reads

\beq
[ a^{i} x^{i} , b^{j} x^{j} ]_{H} \ = \ \imath \hbar \epsilon^{ijk} 
a^{i} \ b^{j} \  M^{kl}  \ x^{l} .
\eeq
The RHS is evaluated with Casimir $ G = b^{j} x^{j} $

\beq
[ a^{i} x^{i} , \frac{1}{2} x^{k} M^{kl} x^{l} ]_{G} \ = 
\ - \imath \hbar \epsilon^{ijk} a^{i} b^{j} M^{kl} x^{l} .
\eeq
So (7.28) is satisfied.

3) G , H  both Quadratic

\beq
G \ = \ \frac{1}{2} x^{j} Q^{jm} x^{m} \ \ ; 
\ \ \ \ \ \ \ H = \frac{1}{2} x^{k} M^{kl} x^{l} ,
\eeq

with Q , M  both real symmetric matrices. 
By Leibniz's  rule  we consider the rel.(7.28) in the
form

\beq
[ x^{i} , G ]_{H} \ = \ - [ x^{i} , H ]_{G}, \ \ \ \ \ \ \ \ i=1,2,3.
\eeq
We demonstrate  its validity by evaluating separately both sides.
Its LHS gives :

\beq
[ x^{i} , \frac{1}{2} x^{j} Q^{jm} x^{m} ]_{H} = \imath \hbar \epsilon^{ijk}
( Q^{jl} M^{km} + Q^{jm} M^{kl}) x^{l} x^{m}.
\eeq
By exchanging $ Q \leftrightarrow M $ and $ j \leftrightarrow k $ 
we similarly evaluate the RHS.
and get :

\beq
[ x^{i} , \frac{1}{2} x^{k} M^{kl} x^{l} ]_{G} =  \ \imath \hbar \epsilon^{ijk}
(M^{kl}Q^{jm} + M^{km} Q^{jl} ) x^{l} x^{m}.
\eeq
This checks the validity of (7.38). 
In effect this implies that it holds also true

\beq
[ F , G ]_{H} \ = \ - [ F , H ]_{G} ,
\eeq
for the cases G and H being either linear or quadratic with F being any polynomial.
To go one step further we have to consider cases where H is cubic and G is either linear
or quadratic and so on. These cases require the construction of 
non-Linear Lie algebras with cubic Casimir or quadratic right hand side 
( quadratic Lie algebras). 
We defer these considerations to a future work.

The main point of this section is to examine the validity of the 
fundamental identity (FI) under the definition (7.27). This is: 
 
 \beq
 [ [ F , G ]_{H} , K ]_{L} \ = \ [ [ F , K ]_{L} , G ]_{H} +
 [ F , [ G , K ]_{L} ]_{H} + [ F , G ]_ {[ H , K ]_{L}}.
 \eeq

We shall check below the above relation, at the level of linear Lie-algebras . 
We must consider the cases where H and L as well as $ [  H , K ]_{L} $ 
are quadratic polynomials ( for linear it is trivial ) and this implies that 
K must be linear. 

\beq
H = \frac{1}{2} x^{k}M^{kl}x^{l} , \ \ \ \ L = \frac{1}{2} x^{j} Q^{jm} x^{m} , \ \ \ \ 
K = x^{r} , \ \ \ \ F = x^{p} , \ \ \ \ G = x^{q} , \ \ ; \ \ \ k,l,j,m,p,q,r=1,2,3 ,
\eeq
where M , Q are real symmetric $ 3 \times 3 $ matrices. The FI becomes

\beq
[ [ x^{p} , x^{q} ]_{H} , x^{r} ]_{L} = [ [ x^{p} , x^{r} ]_{L} , x^{q} ]_{H} +
[ x^{p} , [ x^{q} , x^{r} ]_{L} ]_{H} + [ x^{p} , x^{q} ]_{[H , x^{r}]_{L}} .
\eeq  
The Casimirs H , L being quadratic give rise to linear Lie-algebras,

\beq
[ x^{p} , x^{q} ]_{H}     =    i \hbar \epsilon^{pqk} M^{kl} x^{l},  \ \ \ \ \ \ \
 [ x^{p} , x^{r}]_{L}   =  \imath \hbar  \epsilon^{prj} Q^{jl} x^{l} \ \ ; \ \ 
p,q,r,j,l=1,2,3 ,
\eeq
while the third Casimir $ [ H , x^{r} ]_{L} $ has to be evaluated

\beq
[ H , x^{r} ]_{L} \ = \ \frac{1}{2} \ M^{kl} \ [ x^{k} x^{l} , x^{r} ]_{L} \ = \ 
 \frac{\imath \hbar}{2}  M^{kl} Q^{jm} ( \epsilon^{lrj} x^{k}x^{m} + \epsilon^{krj} 
x^{m} x^{l} ).
\eeq
There are three terms of similar nature in (7.44), the LHS and the first two in the RHS , 
which we label as  RHS1 and RHS2 . They are given as follows:

\beq
LHS = [ [ x^{p} , x^{q} ]_{H} , x^{r} ]_{L}  = 
 -\hbar^{2} \epsilon^{pqk} \epsilon^{lrj} M^{kl} Q^{jm} x^{m} , 
\eeq

\beq
RHS1  =  [ [ x^{p} , x^{r} ]_{L} , x^{q} ]_{H}  =  -\hbar^{2} \epsilon^{prj} 
\epsilon^{mqk} Q^{jm} M^{kl} x^{l} ,  
\eeq

\beq
RHS2  =  [ x^{p} , [ x^{q} , x^{r} ]_{L} ]_{H}  =  
- \hbar^{2} \epsilon^{qrj} \epsilon^{pmk} Q^{jm} M^{kl} x^{l} .
\eeq
In order to evaluate the third term of the RHS, 
$RHS3$, we rewrite the Casimir rel.(7.46) in a
convenient form :

\beq
[ H , x^{r} ]_{L} \ = \ \imath \hbar  \frac{1}{2} x^{m} G^{ml}_{r} x^{l} \ \ ; \ \ \ \ \ 
r,m,l=1,2,3 ,
\eeq

where $ G^{ml}_{r} $ is a real symmetric 
$ 3 \times 3 $ matrix in the indices m,l, 
$ \forall r=1,2,3  $ :

\beq
G^{ml}_{r} = \epsilon^{krj} \ \ ( \ \ M^{kl} \ Q^{jm} \ + \ M^{km} \ Q^{jl} ).
\eeq
Then 

\beq
RHS3 = [ x^{p} , x^{q} ]_{[H , x^{r} ]_{L}} \ = \ 
- \hbar^{2} \epsilon^{pqm} G^{ml}_{r} x^{l}.
\eeq
By comparing the coefficients of $ x^{l} $, we find that : 

\beq
\epsilon^{pqk} \epsilon^{mrj} M^{km} Q^{jl} = ( \epsilon^{prj} \epsilon^{mqk} + 
\epsilon^{qrj} \epsilon^{pmk} + \epsilon^{pqm} \epsilon^{krj} ) M^{kl} Q^{jm}+
\epsilon^{pqk} \epsilon^{mrj} M^{km} Q^{jl}.
\eeq
As the LHS and the last term in the RHS are equal the parenthesis term must vanish.
By using the identity

\beq
\epsilon^{ijk} = \frac{1}{2} ( i-j)(j-k)(k-i) \ \ ; \ \ \ \ \ i,j,k=1,2,3 ,
\eeq
we find that 

\beq
\epsilon^{prj} \epsilon^{mqk} + \epsilon^{qrj} \epsilon^{pmk} + \epsilon^{pqm}
\epsilon^{krj} = \frac{1}{4} (j-m)(k-p)(k-q)(p-q)(j-r)(m-r)
\eeq
This expression is antisymmetric in j,m and the subsequent summation with the symmetric
matrix $ Q^{jm} $  gives the desired result. 

We proceed to discuss the quantization of the $ T^{3} $  Nambu-Poisson 3-algebra in
rel.(3.20) \cite{Hop2,AF}

\beq
\{ e_{n} , e_{m} , e_{l} \} = -i n \cdot ( m \times l ) e_{n+m+l},
\eeq

where $ ( e_{n} )_{n} \ \in \  Z^{3} $ is the plane wave basis in $ T^{3} $ 

\beq
e_{n}(x) = e^{i n \cdot x} \ \ ; \ \ \ \ \  x \in R^{3} , \ \  n \in Z^{3}.
\eeq
We start with the non-commutative torus algebra given a fixed
 $ l=(l_{1} , l_{2} , l_{3} )
\in Z^{3} $ 

\beq
[ x^{i} , x^{j} ] \ = \ \imath \hbar \epsilon^{ijk} l_{k}.
\eeq
By using the Baker-Cambell-Hausdorf formula for the set of 
exponential operators( 3-d magnetic translations ) 

\beq
T_{n} = e^{i n \cdot x } \ \ ; \ \ \ \ n \in Z^{3} ,
\eeq
we obtain 

\beq
T_{n} T_{m} \ = \ e^{- \frac{\imath \hbar }{2} det (n,m,l)} \ \ T_{n+m} ,
\eeq
or equivalently the Lie algebra of 3-dim. magnetic translations

\beq
[ T_{n} , T_{m} ] \ = \ - 2 \imath \ \mbox{sin} [ \frac{\hbar }{2} 
 \mbox{det}(n,m,l)] \ \ T_{n+m}.
\eeq
This is a generalization of the trigonometric algebra 
in 2-dim. phase space \cite{FFZ}.

Fixing the vector $ l \ \in \ Z^{3} $ we have chosen a Casimir for the algebra 
 (7.56) of a  2-d classical torus $ T^{2} $ embedded in $ T^{3} $ . 
The  $T^{2}$ Nambu-Poisson algebra is :

\beq
\{  e_{n} , e_{m} \}_{e_{l}} \ = \ -\imath n ( m \times l ) e_{n+m} \cdot e_{l}.
\eeq

So $ e_{l}(x) $ is a phase on this surface :

\beq
e_{l}(x) \ = \ e^{ic}.
\eeq
At the quantum level the commutation relation (7.60) should get a phase factor for the 
quantum Casimir 

\beq
[ T_{n} , T_{m} ]_{T_{l}} \ = \ -2 \imath \ \mbox{sin} [ \frac{\hbar }{2}
 \mbox{det}(n , m ,l)] 
T_{n+m+l},
\eeq

\beq
T_{l} \ = \ e^{\imath l \cdot x } = e ^{i c \cdot I}.
\eeq

This means that according to our prescription rel.(7.27) we have 
the quantum 3-torus algebra

\beq
[ T_{n} , T_{m} , T_{l} ] \ = \ - 2 \imath \ \mbox{sin} [ \frac{\hbar }{2} 
\mbox{det}(n,m,l)] T_{n+m+l},
\eeq

as a foliation of the algebra (7.61) for all values of $ l  \ \in \ Z^{3} $ or 
of the Casimir

\beq
l \cdot x = c \cdot I.
\eeq
We close this last section by discussing the case of $ S^{3} $  quantum  3-algebra. 
We choose four quantum coordinates $ x^{i}, \ \ i=1,2,3,4 $ 
satisfying the commutation
relations 

\beq
[ x^{i} , x^{j} ] = \imath \hbar  \ \epsilon^{ijkl} \  \alpha^{k} \  x^{l} , 
\ \ \ \ \ \ i,j,k,l=1,2,3,4
\eeq
where we have two Casimirs 

\beq
C_{1} = \alpha \cdot x \ \ ; \ \ \ \ \ \ \ \alpha \in R^{4} ,
\eeq
a quantum $ R^{3} $ space embedded in $ R^{4} $ and 

\beq
C_{2} = \frac{1}{2} x^{2}. 
\eeq
The algebra (7.67) is an elegant way to write the little group subalgebra fixing a
four vector $ \alpha $ of $ SO(4) $ which is an $ SO(3)$ .

If the values of the Casimir $ C_{1} $ belong to the range 

\beq
- \sqrt{2 C_{2}} < C_{1} < \sqrt{2 C_{2}},
\eeq
the $ R^{3}$  quantum space intersects the quantum sphere $ S^{3} $ into an $ S^{2} $ 
quantum sphere of radius $ \sqrt{2C_{2} - C_{1}^{2}} $ . 
So we can obtain the quantum 
$ S^{3} $ sphere as a foliation of quantum $ S^{2} $ spheres analogous to the 
classical case. 

We proceed to define the quantum $ S^{3} $ 3-bracket as follows:

\beq
[ x^{i} , x^{j} , x^{k} ]_{S^{3}} \ = \ [ x^{i} , x^{j} ]_{x^{k} , C_{2} } \ \ ; 
\ \ \ \ \ \ 
i,j,k=1,2,3,4 .
\eeq
This means that we have chosen $ \alpha^{i} = \delta^{ik} $
Hence we obtain 

\beq
[ x^{i} , x^{j} , x^{k} ]_{S^{3}} \ = \ \imath \hbar \epsilon^{ijkl} x^{l} 
\ \ ; \ \ \ \ \ \ i,j,k,l = 1,2,3,4 .
\eeq
The quantum 3-algebra (7.73) satisfies the fundamental identity since its structure
constants are identical to the corresponding classical Nambu-Poisson 3-algebra.  
In our case the
validity of the Leibniz property is obvious for the first two arguments. 
According to this construction the quantization can be carried out for  any quadratic 
3-manifold embedded in $ R^{4} $. 

We close this last section with some comments. 
Our proposal is primarily guided by the 
consistency of the quantum Nambu-Heisenberg evolution equations
as well as  for their uniqueness in time evolution. 
Equally important is the validity of 
the quantum Liouville equation in a three dimensional phase space(PS).
This leads to the following picture which emerges from the last two sections.

The quantum three dimensional phase space, is a foliation of two
dimensional quantum phase spaces, which is  parametrized by the value 
of the phase space defining Casimir. The choise of the second dynamical 
Hamiltonian can be arbitrary and the algebra of the three  quantum coordinates 
is preserved in time. If we want to change the roles of the 
two Hamiltonians,
then for linear or quadratic ones we checked that this is equivalent with time reversal.
This approach uniquely determines  the quantum Nambu  3-brackets. In 
the last section
we demonstrated that the resulting quantum Nambu-Lie 3-algebras can consistently be 
defined for all three  spaces $R^3, S^3, T^3$ as well as for 
quadratic three dimensional 
manifolds embedded in $R^4$. We will come back with explicit constructions of 
representations of the above quantum NL 3-algebras\cite{AFN2}.

%----------------------------------------------------------------------------------------

\section{ Conclusions-Open Problems}

%----------------------------------------------------------------------------------------

In this work we presented a geometrical perspective for classical and quantum Nambu dynamics in three 
dimensional phase space manifolds. 
The two Hamiltonians are interpreted, the first one as the one who defines 
the two dim  phase space geometry, embedded in the 3-d
phase space, while the second one gives the dynamics of the  trajectories on 
the 2-d phase space. 
This view persists in all higher n-dimensions of phase space where there exists 
n-1 Hamiltonians. 
We choose n-2 of them to define a 2-d phase space embedded in 
n-dimensions with the (n-1)th Hamiltonian to define the trajectories.

This perspective stressed, in effect, the importance of the 
$ SDiff({\cal M}_{3}) $ group as the
all embracing framework of possible Nambu 3-d Hamiltonian systems which, after all, 
are the
flow equations for stationary incompressible fluids in the manifold. 
We presented explicit constructions, in the Clebsch-Monge gauge, 
of the structure constants of the Nambu-Poisson 3-algebras for the cases of $R^{3}$, 
the torus $T^{3} $ 
and the sphere $ S^{3} $ as well as of quadratic 3-d manifolds embedded in $ R^{4} $ . 
The
foliation of the three dimensional phase space by arbitrary two 
dimensional symplectic manifolds, whose quantization is well known either 
by operator methods or $\star $ -quantization techniques
(path integral methods), motivates the definition of the quantum 3-bracket (or 3-geometry)
as a foliation of quantum 2-brackets (commutators).

The Nambu 3-bracket is a volume density element defined by three smooth functions on
$ ({\cal M }_{3})$ which defines intersecting surfaces. Systems of triply orthogonal
surfaces on $ R^{3}$ space have interesting applications in hydrodynamics, 
in integrable potentials in Quantum mechanics as well as in Soliton theory. 
There are corresponding non-linear Lie algebras which appear as symmetries 
of such dynamical systems( $ W_{3}$ algebras , quantum groups , etc  ).
Our aproach has obvious connections with the general framework of 
non-commutative geometry.

The quantum 3-commutator should be viewed as the corresponding quantum volume density element. 
It is associated, in our case, with the intersection of quantum (fuzzy) surfaces.
We believe that quantum 3-algebras ( constant, linear  or generally non-linear)
 is a new interesting area of mathematics in itself, with importance as well 
for the quantization
of fluid dynamics and more generally for the geometry of 3-d 
manifolds(branes) such as our physical space (quantum gravity). 
Interesting open questions are the construction of a consistent matrix model for
interacting multiple $M_{2} $ branes , a Matrix model for light cone 3-branes and finally
matrix quantization of Euler fluid dynamics including Vortices and Turbulence.

%---------------------------------------------------------------------------------------

\section{Acknowledgements}

%---------------------------------------------------------------------------------------

For discussions we thank C.~Bachas, I.~Bakas, J.~Hoppe, J.~Iliopoulos, A.~Kehagias, C.~Kokorelis, 
S.~Nicolis, A.~Petkou, G.~Savvidy, S.Sheich-Jabbari and C.~K.~Zachos. 
M.A. and E.F. acknowledge partial support from the E.U. networks : 
UniverseNet  MRIN-CT-035863 as well as of the  MRTN-CT-2004-512194-503369 .

%-----------------------------------------------------------------------------------------

\end{document}
  